\documentclass{aastex63}

\usepackage{xfrac}
\usepackage{float}
\usepackage{physics}
\usepackage{amssymb}
\usepackage{savesym}
\savesymbol{tablenum}
\usepackage{rotating}
\usepackage[section]{placeins}
\usepackage[noabbrev]{cleveref}
\usepackage{subfigure}
\usepackage{gensymb}
\usepackage{siunitx}

\newcommand{\HI}{{H}{I}}

\newcommand{\kms}{km s$^{-1}$}
\newcommand{\msun}{\textit{M}\textsubscript{\(\odot\)}} %{$M_{\sun}$}
\newcommand{\dunit}{\textit{m} \textsuperscript{-3}}
\restoresymbol{SIX}{tablenum}
\newcommand{\tus}[1]{$_{\text{2}}$}
\newcommand{\hh}{H$_2$}

\received{June 1, 2019}
\revised{January 10, 2019}
\accepted{\today}
\submitjournal{ApJ}

\shorttitle{Star Formation in Splash Bridges}
\shortauthors{Yeager and Struck}

\begin{document}

\title{Star Formation in Splash Bridges}

\correspondingauthor{Travis Yeager}
\email{yeagerastro@gmail.com}

\author[0000-0002-2582-0190]{Travis Yeager}
\affiliation{Lawrence Livermore National Laboratory \\
7000 East Ave, Livermore \\
Livermore, CA 94550, USA}

\author[0000-0002-6490-2156]{Curtis Struck}
\affiliation{Iowa State University \\
Physics Hall, 2323 Osborn Dr \#12  \\
Ames, IA 50011, USA}

\begin{abstract}

Splash bridges are created from the direct collision of two gas-rich disk galaxies.  These direct collisions can eject gas masses on the order of \SI{e10} \msun{} stripped from the stellar disks of each galaxy.  The Taffy Galaxy system (UGC 1294/5) is a prototypical example of a splash bridge system.  CO observations of the Taffy revealed that its splash bridge contains a mass of \hh{} equal to that of the Milky Way's \hh{} mass. However, the little visible star formation occurring within the bridge highlights the need for models of direct gas-rich disk collisions.  The Arp 194 system displays what may be another splash bridge resulting from the collision between two disk galaxies.  The region between the two stellar disks contains bright clumps of active star formation.  We aim to better understand the conditions for star formation in splash bridges by employing a Jeans criterion to determine where gravitational instabilities occur in the shocked and cooling gas of gas-rich disk collisions. The splash bridge results are obtained from our previous work using a sticky particle code and post-processed.  We find that the inclination between the gas disks and the collision velocity with which the gas collides strongly affects the fraction of gas that will become gravitationally unstable.  Low inclinations between gas disks produce starbursts whereas high inclinations result in steady-rate star formation. The offset of the gas disks at impact will determine how many gas elements directly collide but does not strongly affect the resulting star formation.

\end{abstract}


\section{Introduction: Splash Bridges} \label{sec:intro}

\indent Splash bridges are dynamic structures of interstellar medium (ISM) between gas-rich colliding galaxies.  They are created from the direct collision of gas-rich disk galaxies.  Splash bridges are distinctly different from other structures formed during galaxy interactions in that they are not primarily the result of tidal gravitational forces.  They are a relatively rare and transient phenomenon on galaxy scales. However, as we show below and in our previous papers (\citet{yeager19, Yeager_2020MNRAS.492.4892Y}) they can be produced in relatively off-center collisions between disks with a wide range of relative tilts, and persist for more than 100 Myr.  Splash bridges under most conditions are temporary gas structures whose material will re-accrete back onto the nearest colliding galaxy.  When the galaxies finally merge the material that has not already fallen back into one of the parent disks will also be merged.  Before merging, the splash bridges, like other tidal structures, provide a laboratory for studying modes of star formation, and the structure of the ISM shredded in the disk-disk impact.  The violence of the direct collisions that create these splashes generates gas turbulence that is much stronger than that which is produced in purely tidal tails and bridges.  The transient nature of these bridges leads to their rarity, though the collisions that produce them are a defining event in the galaxies' evolution.  The fact that splash bridges offer insight into the evolution of galaxies and the cosmological structures around them motivates their study, despite their rarity.  In this paper we will focus on models of star formation in splash bridges.

\indent There has been extensive observational work done to study the enhancement of star formation in tidal structures in interacting galaxies, beginning with the deep optical studies of Schweizer (see the overview of \citet{1986Sci...231..227S}), and the early survey of \citet{1990AJ.....99..497S}. While the present work will focus specifically on star formation within splash bridges,  we are also interested in comparing this to star formation in other tidal structures. The literature on this topic now includes hundreds of studies of individual systems.

More recent surveys include the work of \citet{2007AJ....133..791S} (also see also \citet{2014AJ....147...60S} and \citet{2016AJ....151...63S}). These authors used Spitzer mid-infrared (mid-IR) data to study a sample of 35 tidally distorted pre-merger interacting galaxy pairs that were selected from the Arp Atlas \citep{1966ApJS...14....1A}.  When comparing these galaxies to mass-normalized star formation rates an enhancement of about two was observed in the interacting pairs.  Up to 10\% of the IR emission, and presumably the star formation, was attributed to tidal structures. The star clusters in these galaxies were studied in ultraviolet UV to mid-IR in \citet{2016AJ....151...63S}, and compared to a non-interacting sample. As in many studies of individual systems, more luminous star clusters were found in the interacting galaxies.  This further substantiates that interactions enhance star formation. The most luminous of these star clusters had a core-halo structure not seen in more typical clusters, indicating the core-halo structure is influenced by interaction history.

\citet{2011ApJ...731...93M} used optical (Hubble Space Telescope) observations to study star cluster populations in a sample of of tidal tails, finding indications of age-dependent cluster formation. \citet{2013ApJ...768..194M} followed up with a neutral hydrogen (HI) study of their tail sample. In this study they found evidence for an HI threshold for star formation, and widespread turbulent kinematics. The latter could not be accounted for by stellar feedback. Instead they suggested a joint buildup of turbulent pressure and star formation to a maximum in tails. The formation of tidal dwarf galaxies in tails has also been studied quite intensively in the last couple of decades; see the reviews of \citet{2012ASSP...28..305D} and \citet{2012MNRAS.419...70K}. Comparable observational surveys of splash bridges are not available, but these results inspire the examination of similar processes in the models below. 

\citet{2014MNRAS.442L..33R} proposed a theoretical framework for interaction-induced star formation based on a very high-resolution simulation of a typical disk-disk galaxy merger from first approach until the disks dissipate their orbital energy and merge. The simulation showed an enhancement of compressive turbulence in the merger process, especially relative to solenoidal turbulence. This compression produced high-density regions which showed enhanced star formation. In most types of galaxy interaction the tidal compressions that presumably drive this turbulence are driven in turn by differential gravitational forces on large scales.

\indent Splash bridge dynamics are driven in the first place by the more-or-less synchronized direct collisions of gas elements across the two gas disks.  Second, as shown in the models below, the dynamics are the result of additional cloud-cloud collisions that occur in the displaced shearing streams pushed into the bridge. Thirdly, still more small-scale collisions result from fallback out of the bridges. While the second and third processes have some counterparts in tidal bridges and tails, the first is unique to splash bridges. Comparison of the processes leading to star formation in these different environments potentially offers a great deal of information about how compressive turbulence and star formation are driven, and how they develop.
 
In recent years there have been a number of simulation studies of star formation in interacting and merging galaxies. These studies largely focus on the net star formation through the merger process, and generally do not examine the history of star formation in tidal structures see e.g., \citet{2008A&A...492...31D}, \citet{2008MNRAS.384..386C}, and \citet{2019MNRAS.485.1320M}. Note that the last study of this list also used the FIRE 2 thermal model to study the changing thermal phase structure of the gas. To date, there have also been very few simulations of the evolving structure and star formation systematics in tidal bridges and tails. This is despite the fact there have been numerous simulations of tails in individual systems, usually comparing to multi-band observations. An exception to these generalizations is the broad model grid of \citet{2016MNRAS.455.1957B}, and references therein), which shed new light on several aspects of tail evolution. Semi-analytic models of the evolution of caustic waves in ring galaxies and in tidal tails were studied in \citet{2010MNRAS.403.1516S} and \citet{2012MNRAS.422.2444S}, respectively.  Another recent exception is the study of star formation using the FIRE simulations of \citet{2020AAS...23518203M}. They found a dependence between the orbital parameters of the collision and the level of star formation enhancement in tidal bridges. We discuss a similar result in models of splash bridges below.

\indent The first splash bridge was discovered in the Taffy Galaxy system (UGC 12914/5) in a wide-view radio continuum by \citet{condon93}.  The discovery of a radio continuum bridge led to further observations of the system across the electromagnetic spectrum, so that the Taffy system is now the best studied example of the phenomenon. For example, the molecular content in the Taffy bridge was observed through CO emission to be between $2-$\SI{9e9} \msun \citet{braine03}.  An additional $5-$\SI{10e9} and $3-$\SI{6e9} \msun{} of molecular mass exists in the disks of UGC 1294/5, respectively. Further CO(1-0) observations using the Berkeley-Illinois-Maryland-Association (BIMA) telescope, revealed a bright region of CO emission just below the disk of UGC-12915 \citep{gao03}.  The CO emission was found to follow the radio continuum across the Taffy bridge, suggesting star formation as the cause of a substantial portion of the radio continuum emission.  
\indent \citet{peterson12}, observed \hh{} emission in the Taffy splash bridge in mid-IR, finding an average surface mass of \SI{5e6} \msun kpc$^{-2}$ and typical excitation temperatures of $150-175$ K in the estimated \SI{4.2e8} \msun{} warm \hh{} gas.  

\indent X-ray observations of the Taffy bridge using the Chandra observatory reveal soft X-ray emission in most of the radio continuum bridge but offset from the dense CO-emitting gas, \citep{appleton15}.  Nine ultraluminous X-ray sources (ULXs) were discovered in the bridge, the brightest of which is located in the prominent CO region below UGC 12915. 

\indent Recent ALMA observations provide evidence that gas in the Taffy system is highly in a highly disturbed state \citet{almataffy2019}.  Analysis shows that the high-velocity component of the bridge gas is consistent with shocks with velocities of 200-300 \kms. Filamentary, ionized gas structures are revealed and the emission from the dense molecular gas shows evidence that it is also in a disturbed state.  Clumps of molecular gas found in filaments of length 1 kpc in length show broad line widths consistent with 80-150 \kms{} shocks.  However, the molecular material found in the filaments shows no signs of current star formation. Prior to the present series of papers the system, and the structure of the gas bridge in particular, was modeled by \citep{vollmer12}.

\section{Method}

\indent This work builds upon our previous work on splash bridges.  For details on our particle code refer to \citet{yeager19} and for a discussion of the splash bridge model grid used in this paper see \citet{Yeager_2020MNRAS.492.4892Y}.

\indent We summarize here our previous method.  Our model consists of two counter-rotating gas-rich disk galaxies that are placed initially some distance apart.  The mass and size of the two disks are modeled after UGC 12914 and UGC 12915.  For brevity we will refer to the model UGC 12914/5 disks as G1 and G2, respectively.  The rotational velocities of G1 and G2 peak at approximately 300 \kms{} and 240 \kms{}.    \Cref{table:modelparameters} lists the parameters we have varied for the collisions we have modeled.  Gravity is implemented with smooth-rigid-moving disk and halo gravity potentials.  Each galaxy feels the potentials of the other.  Gas particles are treated as massless test particles for their equation of motion but they are assigned a temperature and density.  The gas particles are initialized as the cells of a two-dimensional grid that is overlaid on both gas disks.  The cell size for these models is set at 100 pc which is also a proxy for the resolution of these models.  Gas particles experience collisions with other gas particles whose centers are within their initial radius of 50 pc.  The collision between two gas particles conserves momentum and triggers shock waves that propagate back through the particles.  The collision radius of each gas particle remains fixed at 50 pc, though the actual radius of the gas particles is changed by shock and cooling processes.  The shock waves heat and compress the gas particles as determined by Mach-dependent jump conditions.  Gas cooling is turned on in shocked gas particles after the shock wave travels across the diameter of the gas particle (generally a relatively short time).  The gas cooling processes included are hydrogen and helium line cooling, bremsstrahlung, CII fine structure line cooling, dust thermal cooling, adiabatic expansion cooling, and UV-photoheating.

\begin{table}[ht]
\movetableright=.25in
\begin{tabular}{|c|c|c|c|c|c|c|c|c|c|}
\hline
X & Y & Z & $V_X$ & $V_Y$ & $V_Z$ & $V_{impact}$ & X tilt & Y tilt \\
\hline
0 & .5 & -15 & 0 & 0 & 400 & 938 & 0 & 0 \\
0 & .5 & -15 & 0 & 0 & 400 & 884 & 0 & 20 \\
0 & .5 & -15 & 0 & 0 & 400 & 919 & 0 & 45 \\
0 & .5 & -15 & 0 & 0 & 400 & 869 & 0 & 65 \\
0 & .5 & -15 & 0 & 0 & 400 & 934 & 0 & 90 \\
0 & 10 & -15 & 0 & 0 & 400 & 642 & 0 & 0 \\
0 & 10 & -15 & 0 & 0 & 400 & 621 & 0 & 20 \\
0 & 10 & -15 & 0 & 0 & 400 & 638 & 0 & 45 \\
0 & 10 & -15 & 0 & 0 & 400 & 625 & 0 & 65 \\
0 & 10 & -15 & 0 & 0 & 400 & 649 & 0 & 90 \\
0 & 10 & -15 & 0 & 0 & 400 & 654 & 90 & 0 \\
0 & -10.6 & -10.6 & 0 & 282 & 282 & 900 & 0 & 0 \\
0 & 20 & 20 & 0 & -350 & -350 & 1018 & 0 & 0 \\
\hline
\end{tabular}
\caption{Initial Parameters for All of Our Simulated Galaxy Disk-Disk Collisions. The first three columns X, Y, and Z in kpc are the position of G2 relative to G1. $V_{X}$, $V_{Y}$, and $V_{Z}$ are the initial velocities in \kms{} of G2. $V_{impact}$ in \kms, is the magnitude of the net velocity at which the gas disks collide. This does not include the counter-rotation velocities that also contribute to the gas cloud–cloud collision velocities. G1 is initially at rest in all runs. X tilt and Y tilt are angles in degrees at which the gas disk of G2 is inclined relative to the X- and Y-axis, respectively.\label{table:modelparameters}}

\end{table}

\indent Our approach to star formation is similar to that employed by the FIRE group \citep{2018MNRAS.480..800H}.  Specifically, we test clouds against a critical Jeans parameter and then require one freefall time to be completed before assuming star formation begins.  Once the shock has propagated through the cloud, it is assumed to be at the density and temperature given by the post-shock jump conditions throughout.  Cooling is turned on and from that time the mean density and temperature are computed self-consistently at constant pressure.  At the end of every timestep the gas cloud density and temperature are used in \cref{eq:jeanslength} to calculate the Jeans length.  The ISM of our initial disks is modeled as five discrete gas phases, with initial temperatures and densities typical of the phases in a gas-rich Sc galaxy.  Each initial disk is divided into a uniform grid of cells, and each grid cell is assumed to contain a spherical gas element.  Each of these gas elements is treated as a test particle (i.e., no self-gravity) and will be referred to in this paper as a "cloud."  The initial temperature and density of each cloud are assigned by its phase.  The gas cloud density also depends on the gas clouds initial position in the galaxy disk.  Gas clouds of a similar phase will be higher in density at the center of the galaxy disk than at the edges.  The initial distribution of each phase of ISM in the initial disks can have a strong influence on the resulting bridge evolution and structure, since the cloud density will change its inertial resistance to the gas ram pressure in a collision:

\begin{equation} \lambda_{Jeans} = \sqrt{\frac{15 k_{B} T}{4 \pi G \mu m_{h} \rho}}
\label{eq:jeanslength}.
\end{equation}

A value of $\frac{5}{3}$ is taken for the heat capacity ratio $\gamma$, $\mu$ is the average molecular weight in the gas, $\rho$ is the gas cloud mass density, $m_{h}$ is the mass of hydrogen, \textit{G} is the gravitational constant and \textit{T} is the gas temperature.  If a gas cloud's radius exceeds that of its current Jeans length the cloud begins gravitational collapse.  The beginning of gravitational collapse triggers the gas cloud freefall time countdown.  The freefall time is only dependent on the cloud's density ($\rho$),  

\begin{equation} t_{freefall} = \sqrt{\frac{3}{2 \pi G \rho}}
\label{eq:freefalltimescale}.
\end{equation}

Once the freefall time is completed we consider that particle to be a site of new star formation.  Should a gas cloud experience another shock after going gravitationally unstable a new freefall time is calculated from its new temperature and density state.  In the case of multiple concurrent freefall times, star formation is considered to begin at the end of the first completed freefall.  

\indent Gas clouds experience collisions with other gas clouds that are within their radius.  Collisions trigger shock waves through each cloud that will heat the gas and drive up the density.  Shocks will drive much of the ISM in the galaxies into the \SI{e8} \dunit{} and greater density range.  The high shock heating temperatures briefly keep these clouds stable but the gas clouds will further contract increasing their density as they cool.  This leads the cooling, dense gas clouds to quickly reach their Jeans Length, triggering gravitational freefall.  

\indent The initial model disks include a number of distinct gas phases, and the initial phase of \hh{} is gravitationally unstable from the outset, though this phase only accounts for less than $10\%$ of the total gas in each galaxy.  The plots below showing rates of cloud collapse include only clouds that collapse after they have been shocked.  This allows us to see where collision-induced star formation occurs.  For the star formation rate calculations we have completed a set for both: 1) clouds allowed to be initially unstable and 2) for only collision-induced cloud collapse.

\section{Timescales for Cooling and Gravitational Instability}

\begin{figure*}
     \centering
          \includegraphics[width=0.9\textwidth]{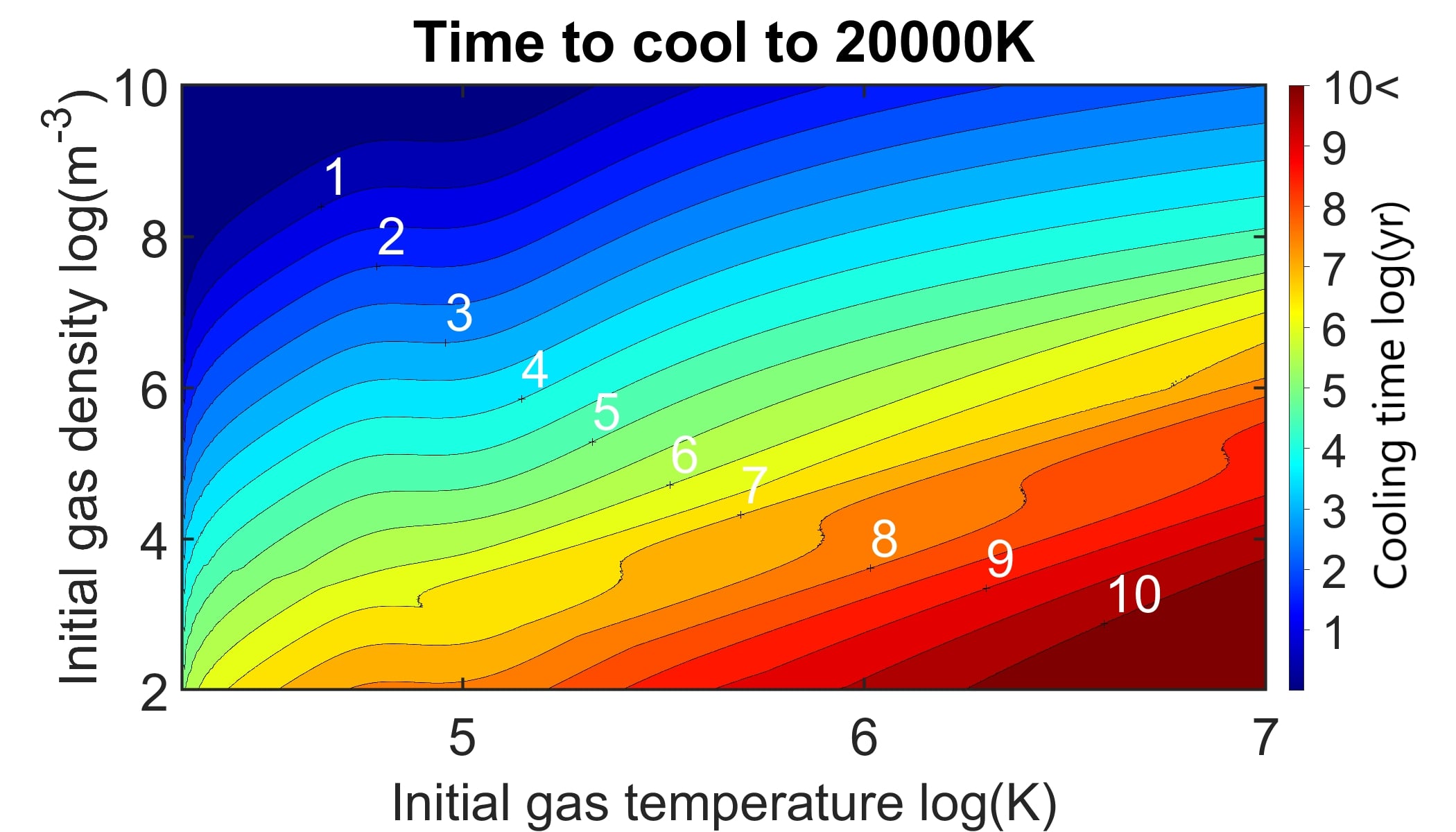}
\caption{Color-coding indicating the time required on a logarithmic scale for a gas cloud with the given initial density and temperature to cool to 20,000 K.  Selected contour line values are labeled in white, with two contour lines per decade.} %
\label{fig:coolingtimetable20000}
\end{figure*}

\begin{figure*}
     \centering
          \includegraphics[width=0.9\textwidth]{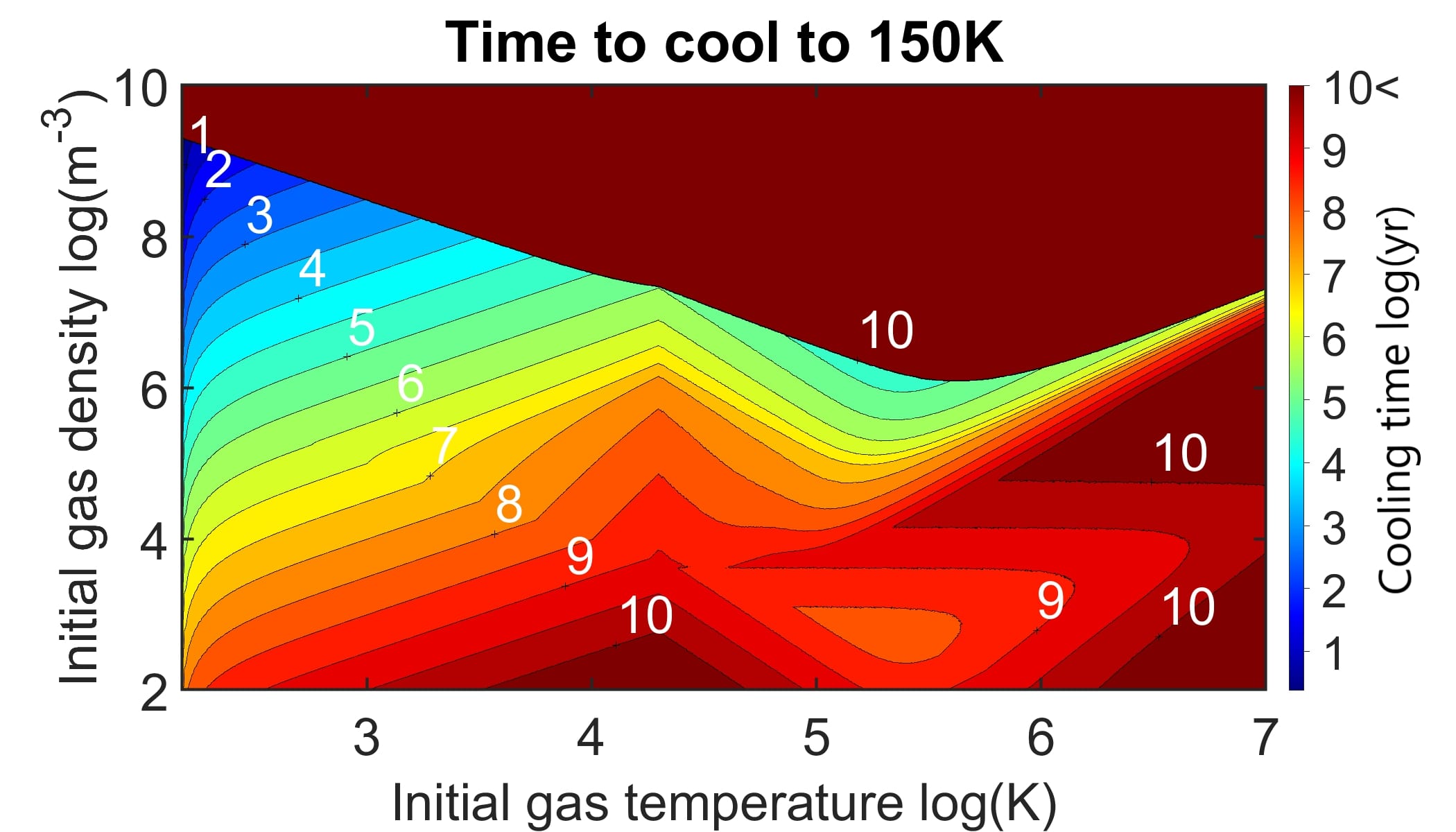}
\caption{Color-coding indicating the time required on a logarithmic scale for a gas cloud with a given initial density and temperature to cool to 150 K.  Selected contour line values are labeled in white, with two contour lines per decade.} %
\label{fig:coolingtimetable150}
\end{figure*}

\indent \Cref{fig:coolingtimetable20000,fig:coolingtimetable150} illustrate the cooling times from our adopted algorithm over a large initial gas temperature and density parameter space.  An upper limit of \SI{e10} yr for the gas to cool is imposed.  Gas elements that do not cool within \SI{e10} yr are assumed not to cool, and the cooling calculation is stopped for them.  In gas clouds whose cooling pathways become weaker than the constant rate of UV photo heating cooling is also halted and assigned a time of \SI{e10} yr.  

\indent In \cref{fig:coolingtimetable150} the \SI{e10} year cooling times at the top of the plot are caused by the gas cloud densities reaching the CII fine-structure critical densities of \SI{2e9} \dunit{}.  For those clouds, the UV heating overcomes the other cooling processes considered so we halt the cooling integration.  These cooling figures provide insight into how the initial velocity of two colliding gas-rich disks can strongly affect the gas cooling time.  In particular, hot gas clouds remain stable against gravitational instability.

\begin{figure*}
     \centering
          \includegraphics[width=0.9\textwidth]{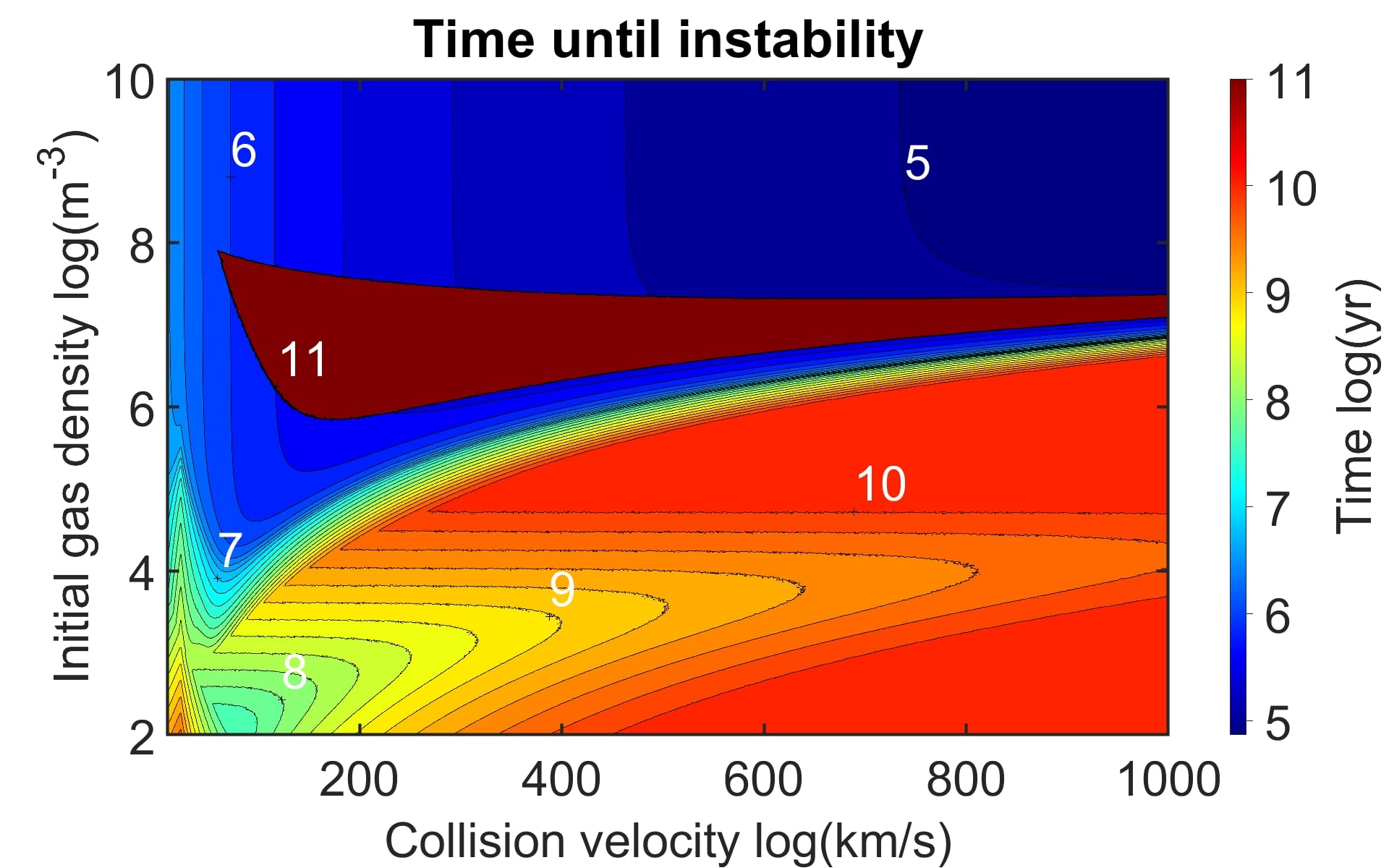}
\caption{Table of gravitational instability timescales for a range of gas cloud densities that undergo direct collisions.  The y-axis covers a wide range initial gas cloud densities.  The x-axis is the range of collision velocities experienced by the gas clouds in a center-of-mass frame.  The color indicates the time taken on a logarithmic scale for the gas to reach the critical Jeans parameter.  Some contour line values are labeled in white.  There are five contour lines per decade.} %
\label{fig:instabilitytimes}
\end{figure*}

\indent \Cref{fig:instabilitytimes} shows calculations done using our shock and cooling algorithm to approximate expected timescales required for the gas in disk-disk collisions to become gravitationally unstable.  This time is the sum of both a single shock crossing time in the cloud and the cooling time taken to reach gravitational instability.  This analysis was done to estimate how the collision velocity between two gas-rich disk galaxies affects the delay to star formation in their subsequent splash bridges.  If two gas-rich galaxy disks collide at 500 \kms{} the gas clouds will still experience a range of collision velocities that depend on the relative densities of the two colliding gas elements.  A range of shock heating strength will also occur.  The strength of shock heating is dependent on the density and local speed of sound in the gas cloud prior to collision.  To narrow the parameter space, for \cref{fig:instabilitytimes} we only include the most abundant neutral \HI{} phase in the analysis.  The \HI{} phase is initialized with a temperature of 5000 K.  The initial temperature affects the speed of sound in the gas cloud, which in turn determines the shock Mach number.  Assuming only gas clouds of similar density collide, then the x-axis center-of-mass collision velocity is equal to the velocity with which the galaxy disks collide.

\indent For the lowest gas densities, the instability time increases quickly with collision velocity.  Instability times decrease as density increases.  The blue and dark blue regions of the figure have instability times of order a million years, and these gas clouds immediately or nearly immediately become gravitationally unstable.  The time delays of \SI{e5}-\SI{e6} yr are the result of the shock crossing times.  

\indent The band of dark red with a time given of \SI{e11} yr in \cref{fig:instabilitytimes} is gas that will never go unstable after a single cloud collision.  Our cooling algorithm is the direct cause of this discontinuity.  With the addition of further cooling pathways, specifically molecular line cooling, for gas of temperatures below 20,000 K and of densities above \SI{2e9} \dunit{} this instability band should disappear.  

\indent The band of stability includes nearly all collision velocities for gas of initial density of \SI{e7} \dunit.  The \HI{} gas initial densities mostly fall in this range.  Gas densities in this range will not decrease their temperature enough by the time they reach the CII cooling critical density of \SI{2e9} \dunit.  However, for the models shown in this paper much of the \HI{} gas does become gravitationally unstable.  This indicates that multiple collisions are able to drive the \HI{} gas out of the region of stability.

\section{Gas Instability in Splash Bridges}

\subsection{Net Rates of Cloud Collapse}
\label{sec:netrates}

\begin{figure*}
  \centering
  \subfigure[]{%
  \includegraphics[width=0.9\textwidth]{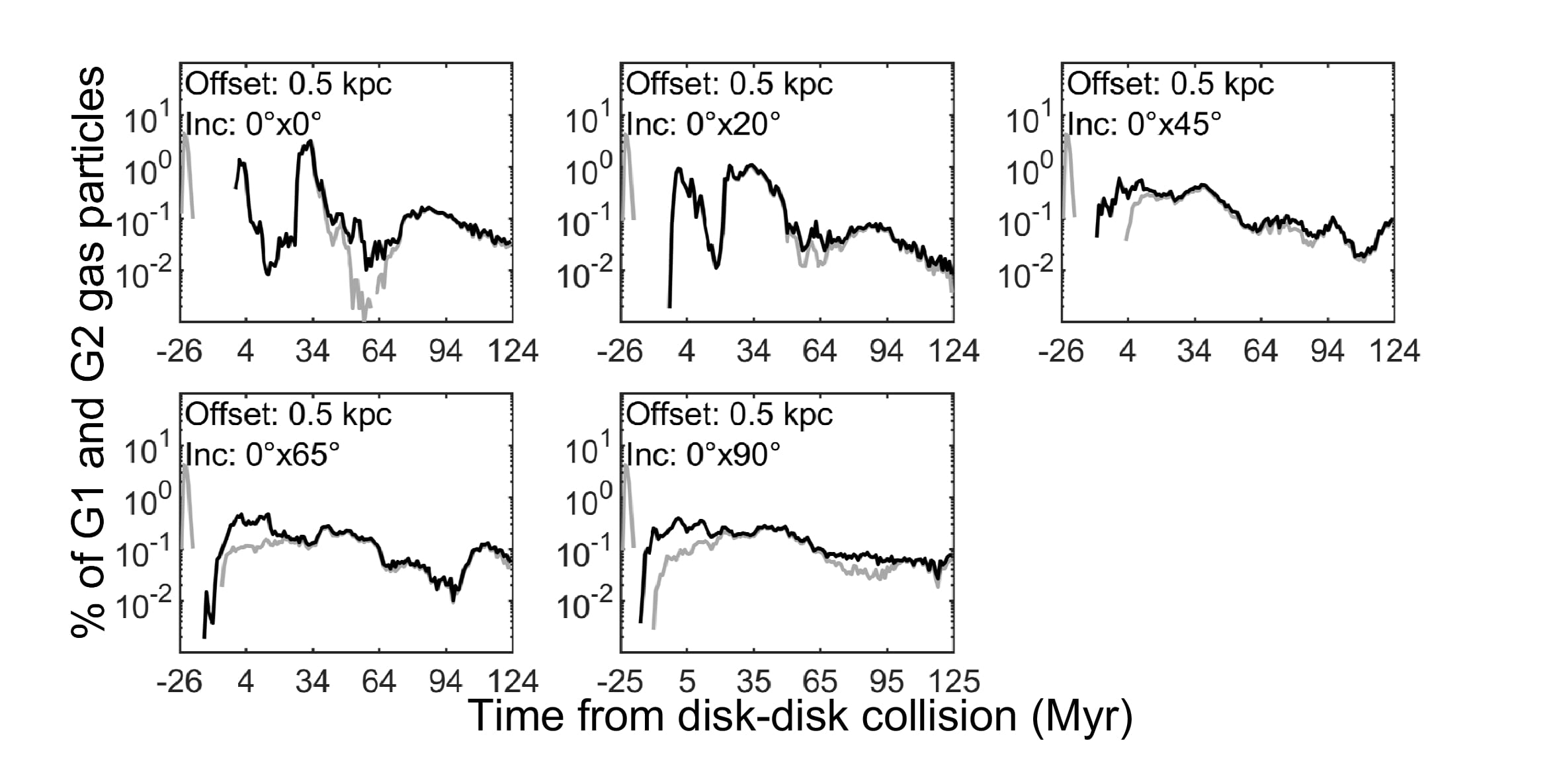}
    }\\
    \subfigure[]{%
  \includegraphics[width=0.9\textwidth]{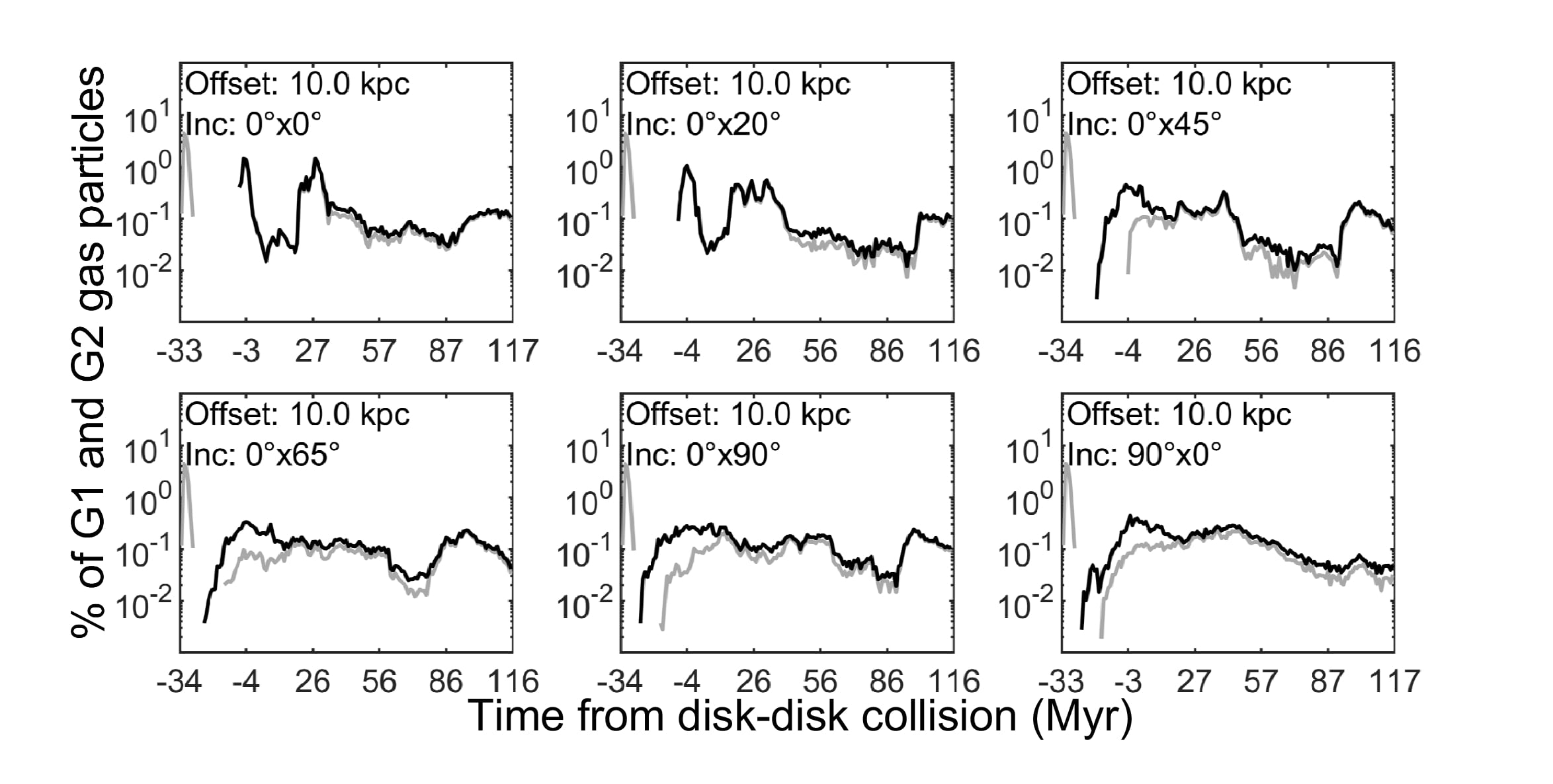}
  }
 \caption{Binned fractional rate of cloud collapses versus time in megayears. Time zero corresponds to when the centers of each galaxy are at their nearest approach. Time bins are 1 Myr in duration.  The black and gray lines represent two different runs.  For the run curve in black, gas clouds are required to experience at least one collision before their collapse can begin, even if the cloud meets the Jeans criteria for collapse.  For the run shown in light gray clouds may begin collapse if the Jeans criteria is met.}%
 \label{fig:sfrate1}
\end{figure*}

\indent Gas cloud collapse rates over time are shown in \cref{fig:sfrate1} for collisions of differing inclination and offset.  The rates at which shocked cloud populations are collapsing are shown with black curves.  Light gray curves show the collapse rate for all unstable clouds.  These differ significantly because our dense \hh{} phase is initially unstable to gravitational collapse.  The star formation rates are  normalized to the total number of gas clouds, which for these runs is 108,221 particles.  Collapse is considered complete once the freefall time has been completed.  The immediate collapse of the \hh{} phase supercedes and so delays the collision-induced cloud collapse after the disk-disk collision by a few megayears in the runs with 45\si{\degree}{} and greater inclinations.

\indent The cloud collapse rates peak at varying times depending on the disk-disk impact parameters.  Low relative disk inclinations ($< 45$ \si{\degree}) have their first peak 2-4 Myr after the disks have collided.  This first peak of gas cloud collapses is then followed by a 30 Myr dark period where very few stars are formed.  A second starburst occurs approximately 30 Myr later.

\indent Collapses begin earlier in the low-inclination collisions with 10 kpc impact offsets.  No gas clouds are seen collapsing before 0 Myr in the low-offset collisions.  The 10 kpc offset collisions result in stronger warping of the galactic disks as they approach one another, so the outer disks make contact before the centers reach their closest approach.  

\indent For runs with relative inclinations of 45\si{\degree} or more the cloud collapse rate is does not peak but grows rapidly before remaining constant for over 50 Myr.  The rates for highly inclined collisions begin before the centers of the disks pass because the inclinations produce prolonged galaxy disk collisions.

\begin{figure*}
  \centering
  \subfigure[]{%
  \includegraphics[width=0.9\textwidth]{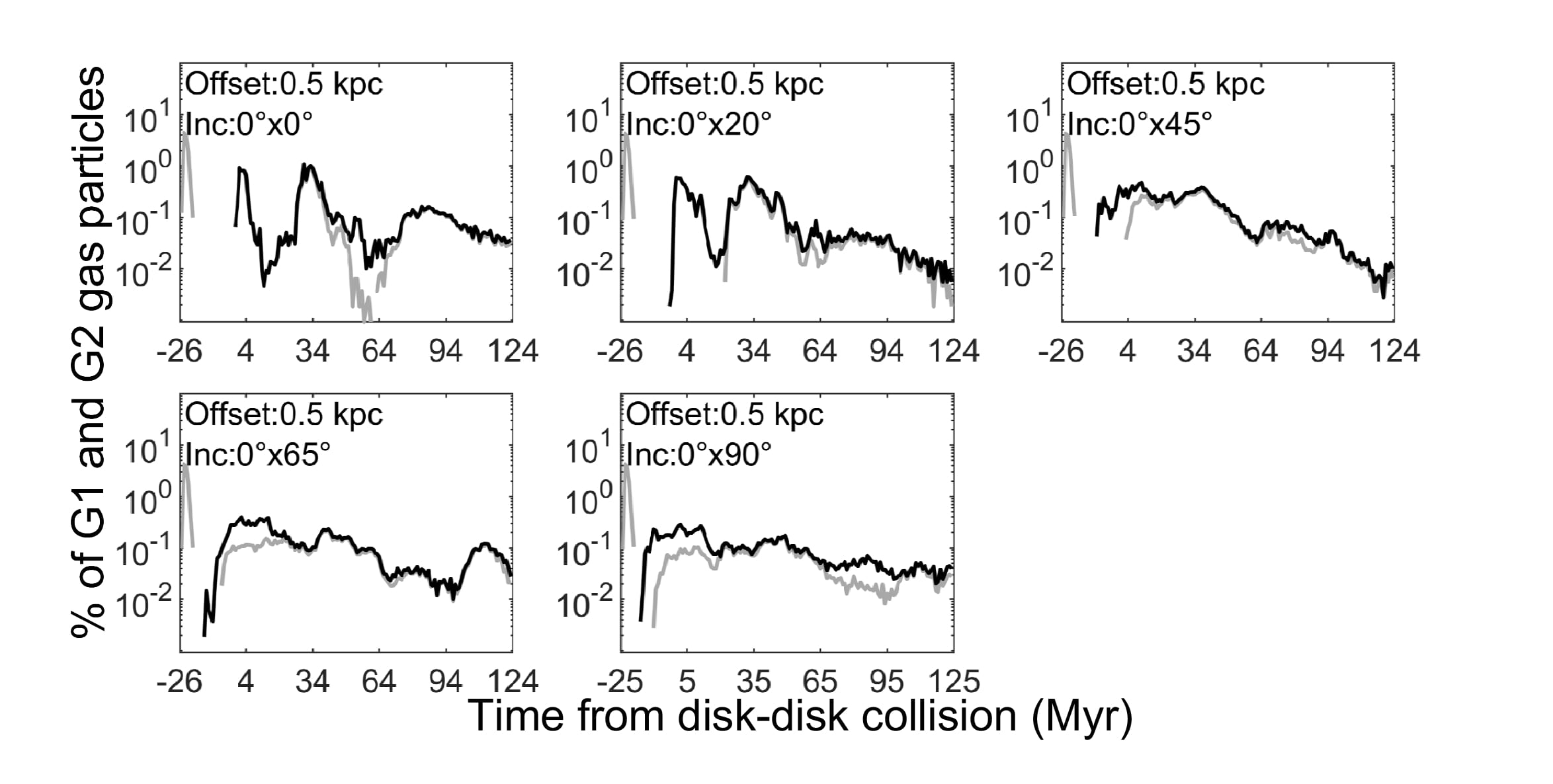}
    }\\
    \subfigure[]{%
  \includegraphics[width=0.9\textwidth]{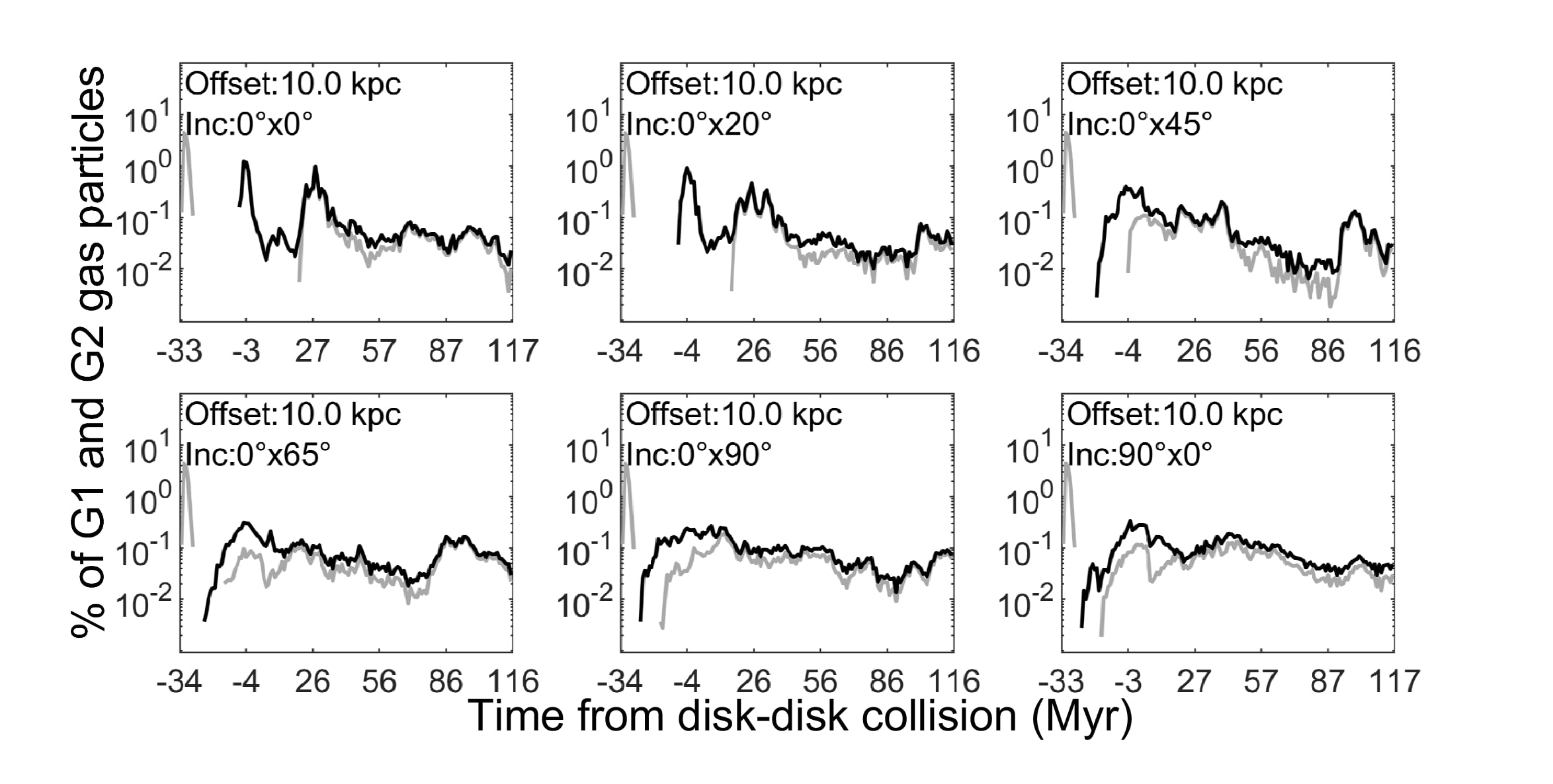}
  }
 \caption{Binned fractional cloud collapse rates for clouds within or near both disks vs. time in megayears for various runs. For further information refer to \cref{fig:sfrate1}.}%
 \label{fig:sfrate1disks}
\end{figure*}

\begin{figure*}
  \centering
  \subfigure[]{%
  \includegraphics[width=0.9\textwidth]{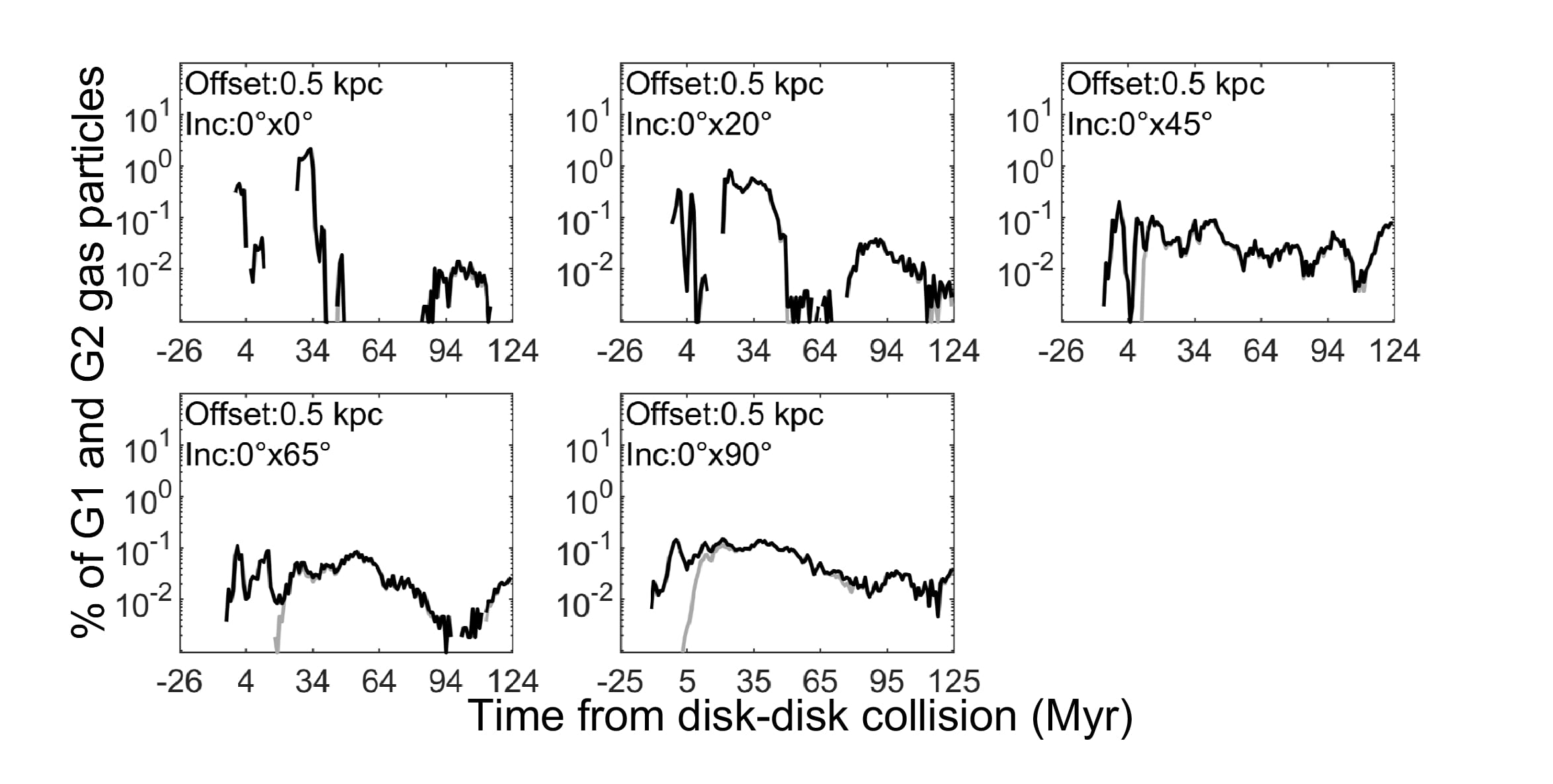}
    }\\
    \subfigure[]{%
  \includegraphics[width=0.9\textwidth]{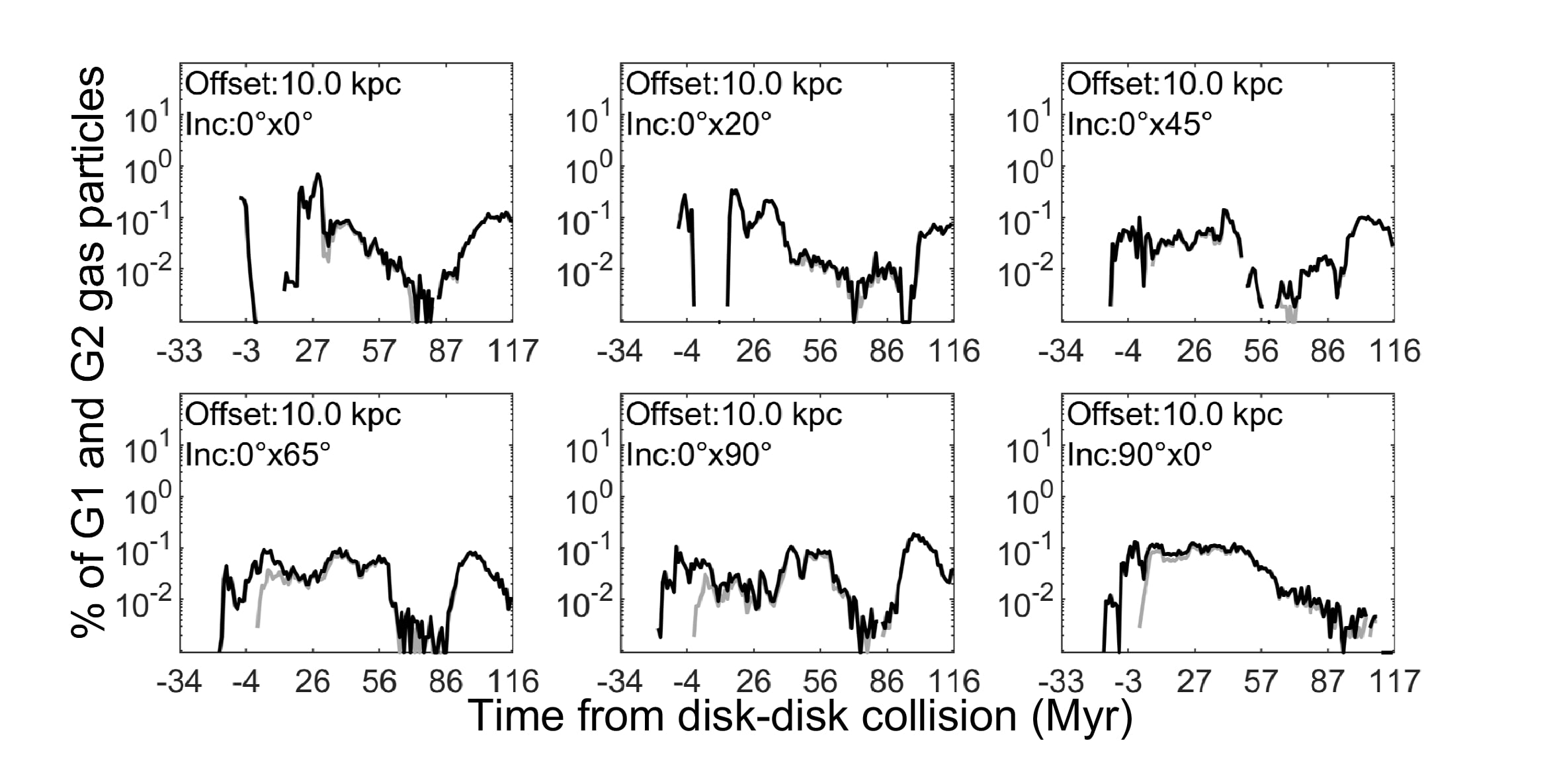}
  }
 \caption{Binned fractional cloud collapse rates for clouds within the bridge (as defined in the text) vs. time in megayears for various runs. For further information refer to \cref{fig:sfrate1}.}%
 \label{fig:sfrate1bridges}
\end{figure*}

\indent \Cref{fig:sfrate1disks,fig:sfrate1bridges} show the cloud collapse rates after dividing the colliding systems according to  some simple cuts in 3D space to separate the bridge gas particles from disk particles. (We henceforth equate cloud collapse rates to relative star formation rates.)  To be considered a part of the bridge a particle must first be at least 1 kpc above G1 in the z-direction.  To handle the full range of possible inclinations of G2 disks we require bridge particles to be outside a cylindrical volume centered on the disk of G2.  The cylindrical region has a radius of $.8*cos(\theta)*D/2) + 1 kpc $ where $D$ is the diameter of G2.  The gas particles at z distance to the center of G2 must also exceed half of the cylindrical height of $.8*cos(\theta)*D/2) + 1 kpc $.  A different rule is used for G2 since it is the slightly smaller and inclined galaxy.  We have tried a cut based off the initial plane of G2’s disk, though because of precession and other tidal effects the cylindrical cut proved more consistent over time.  A gas particle that fails any of these criteria is then considered within a disk for this analysis.  These cuts are made for every timestep in the simulations.  The number of bridge and disk particles varies drastically over the 150 Myr simulation time.  

\indent For low-inclination collisions the gas disks see a factor of 3-6 increase in star formation over the average rate.  The heightened star formation takes place for approximately 20 Myr and begins 20-30 Myr after the disk-disk collision.  Increasing the relative inclination between the galaxy disks quickly washes out the initial peak in disk star formation.  High inclinations see no star bursts in the gas disks but they do maintain a nearly constant star formation rate of strength similar to the rates outside the starburst peaks of low-inclination collisions.  

\indent The star formation histories in the bridges include starbursts which occur approximately 30 Myr after low-inclination disk-disk collisions.  These bursts are about an order of magnitude stronger than those seen in the disks, but in the case of the face-on collision are much shorter lived.  The burst of star formation in the bridge occurs across less than 5 Myr, where as the heightened rates seen in thee disk last on the order of 20 Myr.  Quenching is much stronger in the bridges as well.  In a few of the collisions we can see the disk rates nearing zero for a time, whereas the bridges' ability to form stars plummets to zero at several different times.  The bridges of high-inclination collisions are better at sustaining star formation rates and have no starbursts at any time following the disk-disk collision.  

\subsection{Spatial distribution of collapsing clouds}

\begin{figure*}
     \centering
                \subfigure[]{%
          \includegraphics[width=0.45\textwidth]{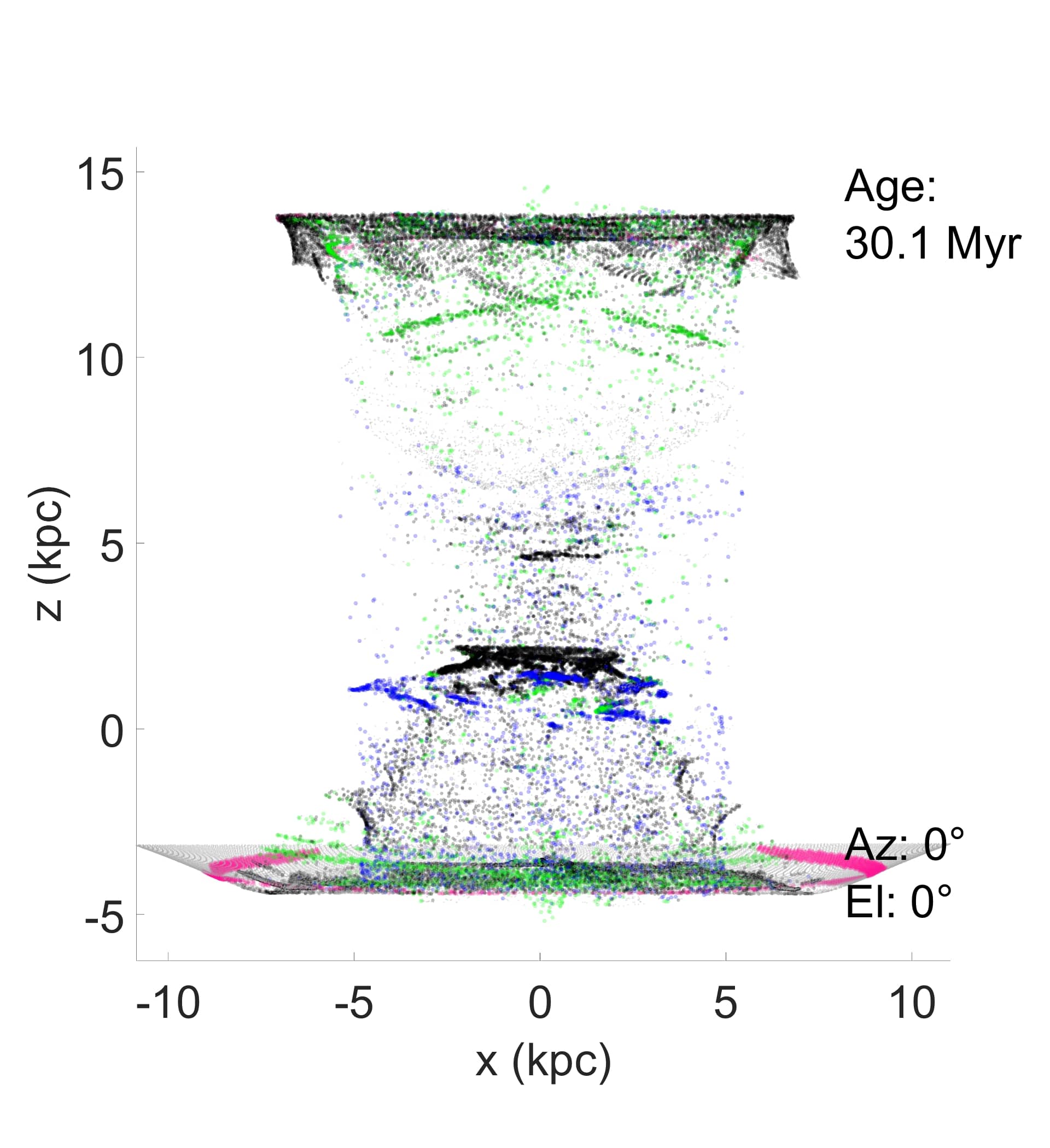}
        }
                \subfigure[]{%
          \includegraphics[width=0.45\textwidth]{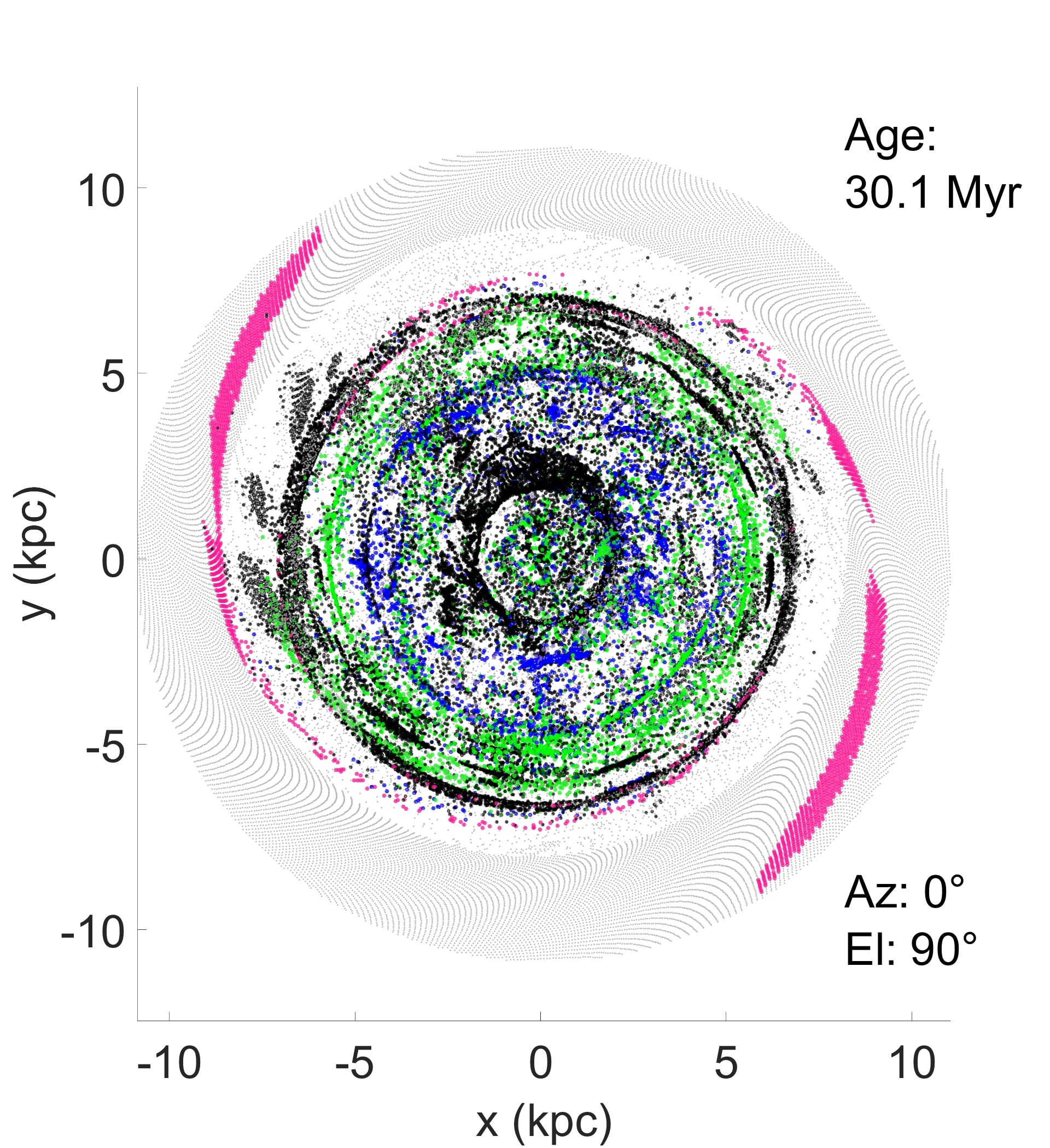}
        }\\ %  ------- End of the first row ----------------------%
                \subfigure[]{%
          \includegraphics[width=0.45\textwidth]{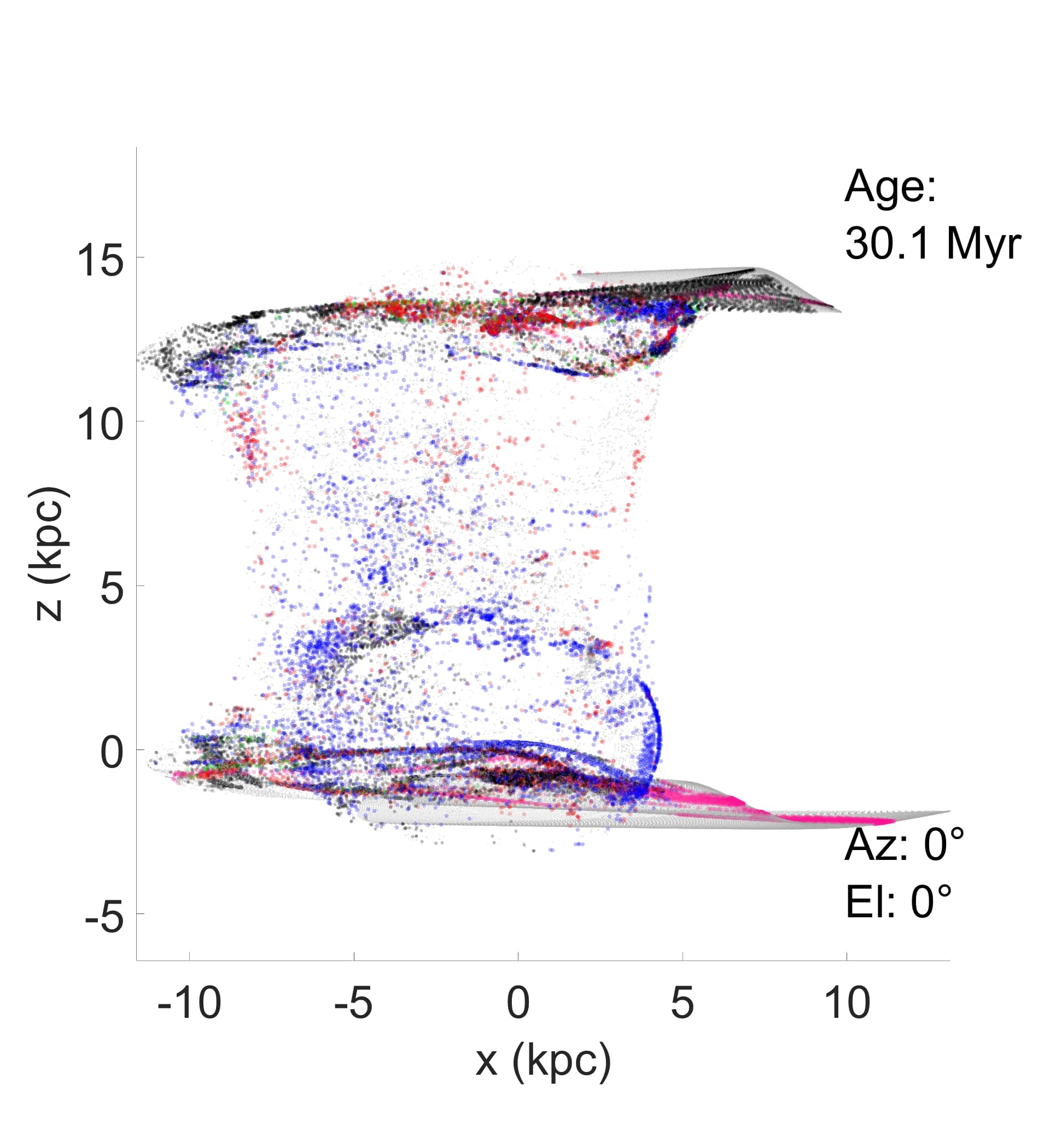}
        }
                \subfigure[]{%
          \includegraphics[width=0.45\textwidth]{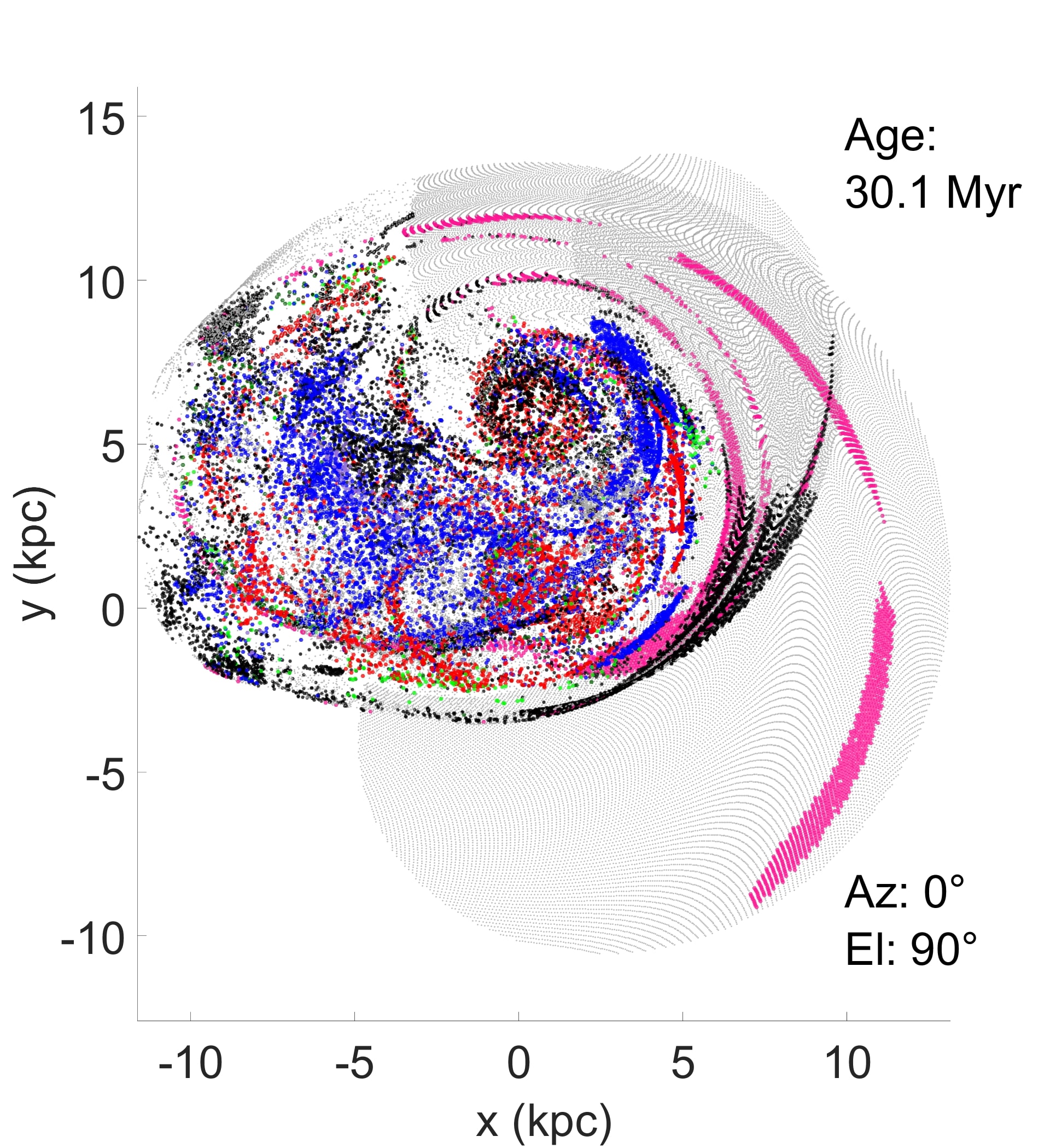}
        }\\ %  ------- End of the second row ----------------------%
\caption{Collisions with relative disk inclinations of $0\si{\degree}$.  Panels (a) and (b) are 500 pc offset collisions.  Panels (c) and (d) are 10 kpc offset collisions.  Color indicates when gas clouds finish collapse.  Pink: $>$10 Myr before disk-disk collision; red: within 10 Myr before disk-disk collision.  Bright green: 0 to 10; dark green: 10 to 20; blue: 20 to 30; purple: 30 to 120 Myr (end of run) after the disk-disk collision.  Black dots indicate gas clouds that are unstable but have not completed a freefall by $t = 120 Myr$.  Light gray dots (barely visible) are gas clouds that never become gravitationally unstable.} %
\label{fig:sf0tilt1}
\end{figure*}

\begin{figure*}
     \centering
                \subfigure[]{%
          \includegraphics[width=0.45\textwidth]{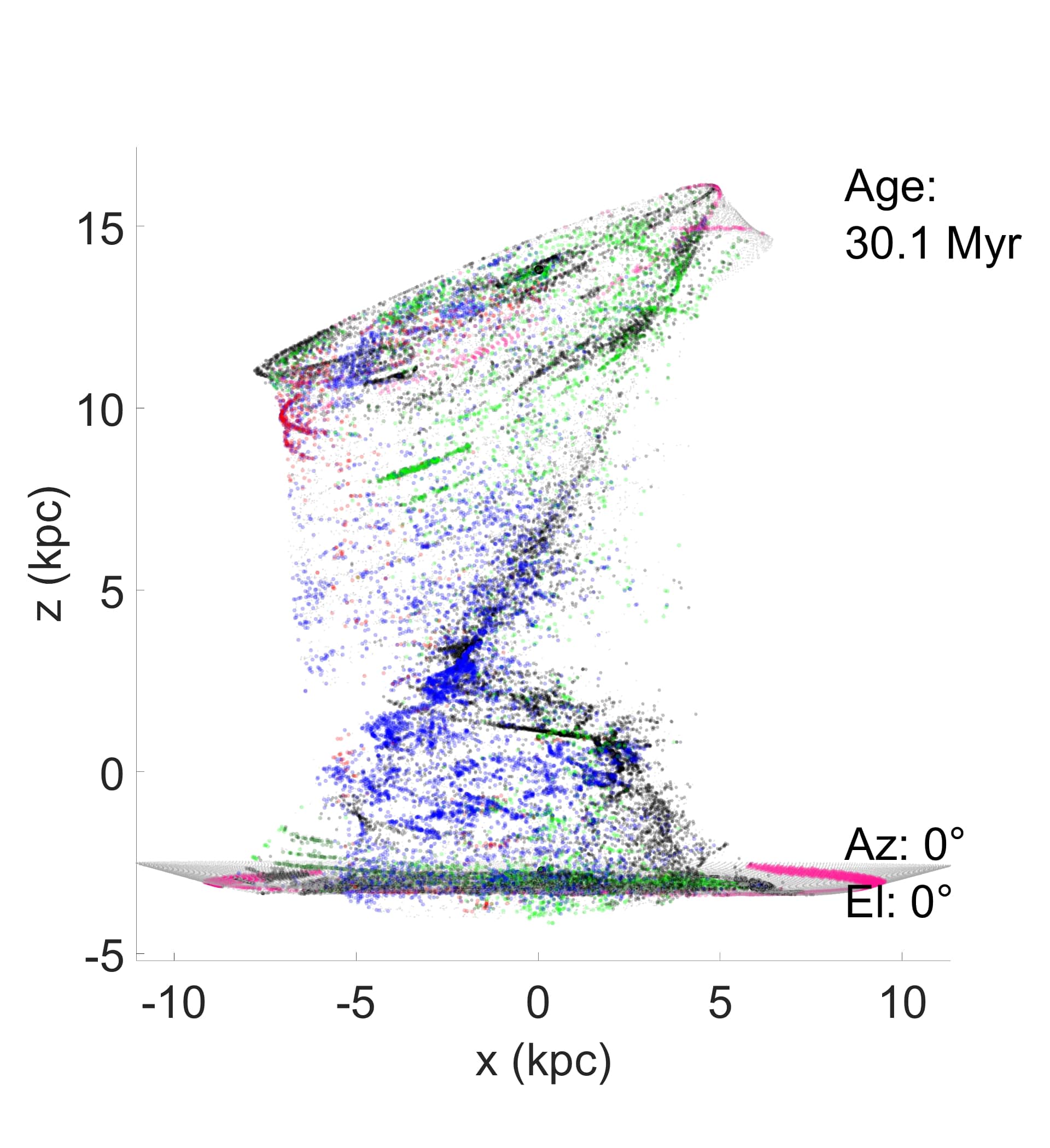}
        }
                \subfigure[]{%
          \includegraphics[width=0.45\textwidth]{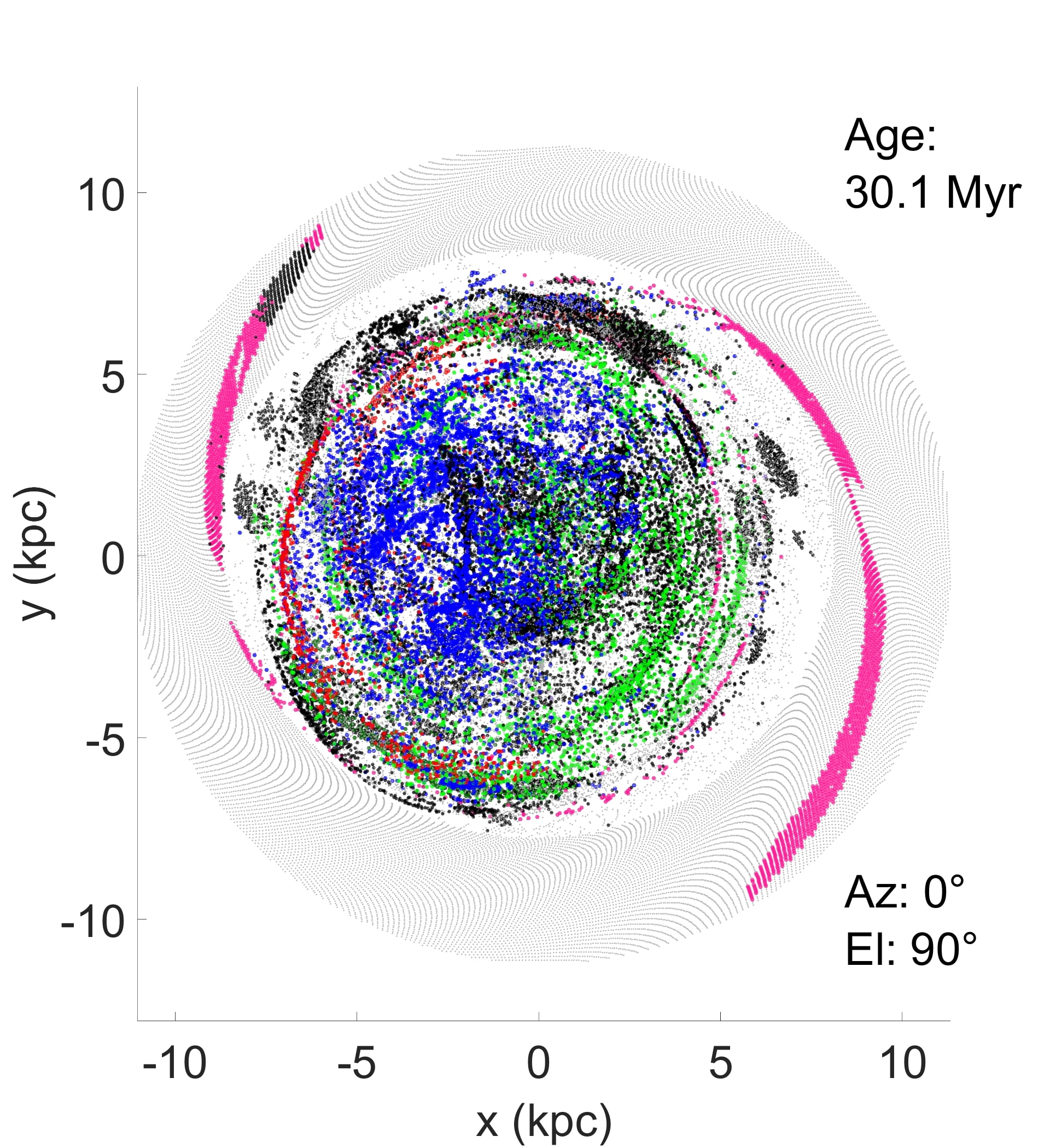}
        }\\ %  ------- End of the first row ----------------------%
                \subfigure[]{%
          \includegraphics[width=0.45\textwidth]{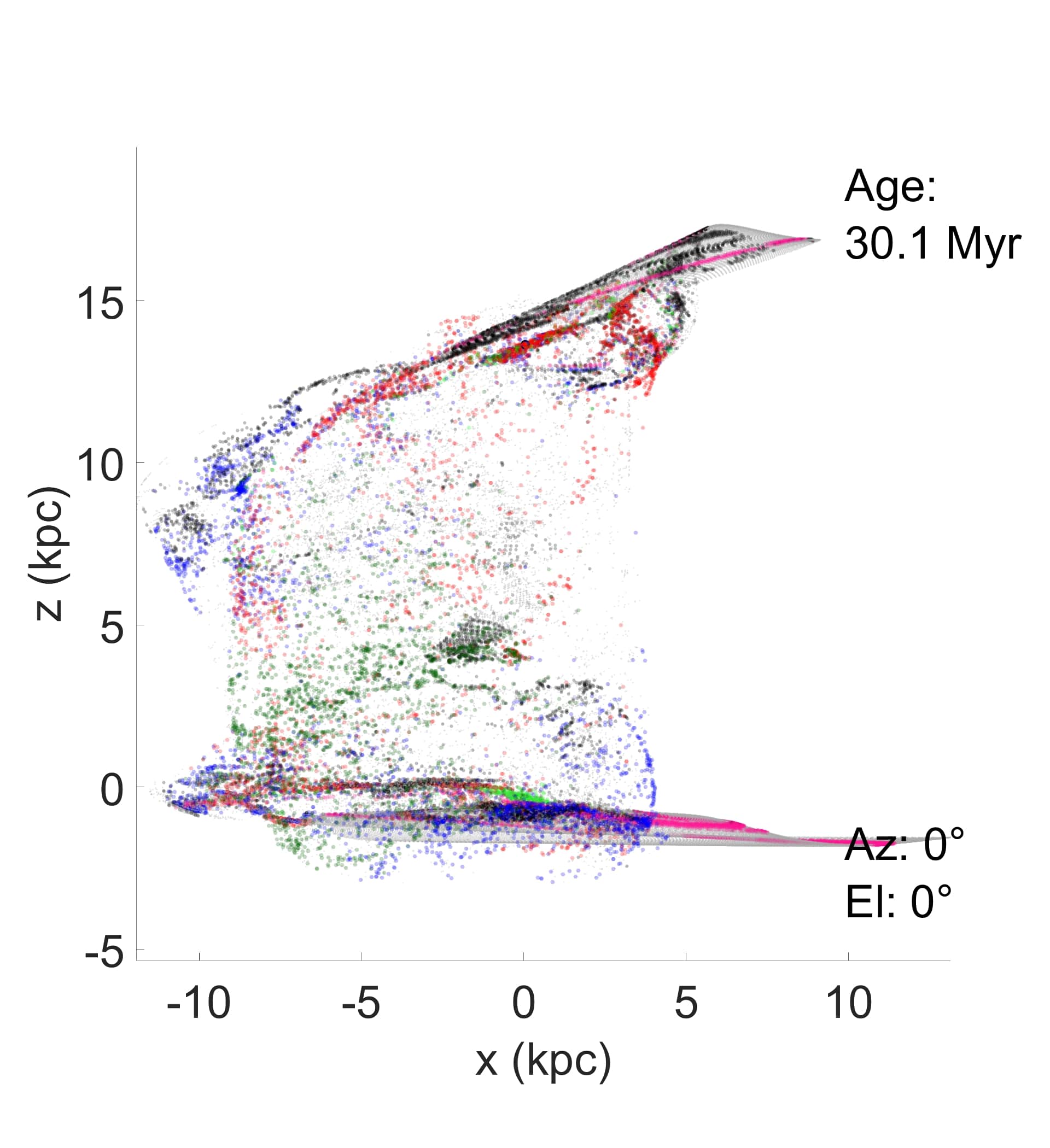}
        }
                \subfigure[]{%
          \includegraphics[width=0.45\textwidth]{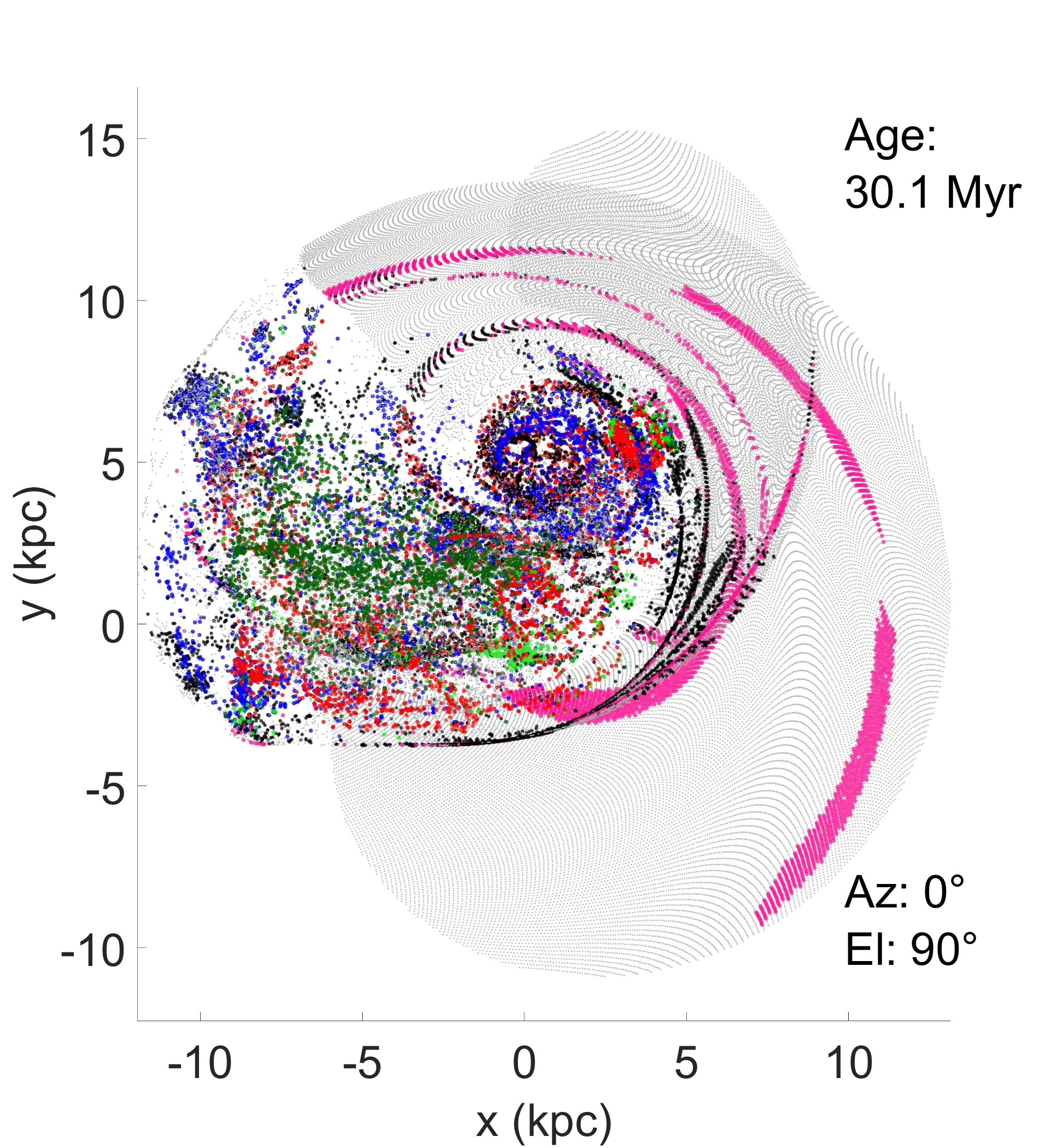}
        }\\ %  ------- End of the second row ----------------------%
\caption{Collisions with relative disk inclinations of $20\si{\degree}$.  Panels (a) and (b) are 500 pc offset collisions.  Panels (c) and (d) are 10 kpc offset collisions.  Color indicates when clouds finish collapse, and is explained in \cref{fig:sf0tilt1}.} %
\label{fig:sf20tilt1}
\end{figure*}

\begin{figure*}
     \centering
                \subfigure[]{%
          \includegraphics[width=0.45\textwidth]{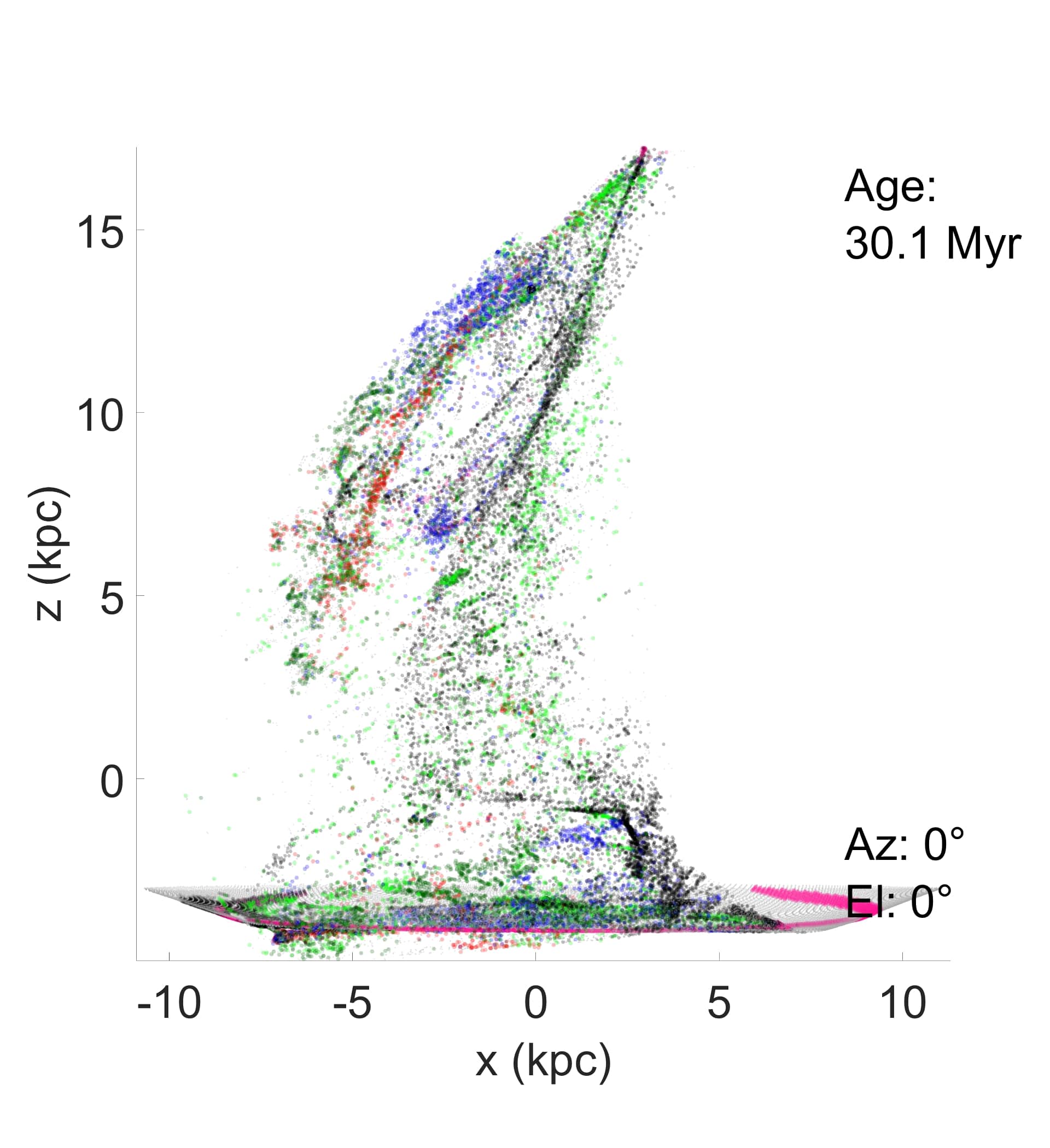}
        }
                \subfigure[]{%
          \includegraphics[width=0.45\textwidth]{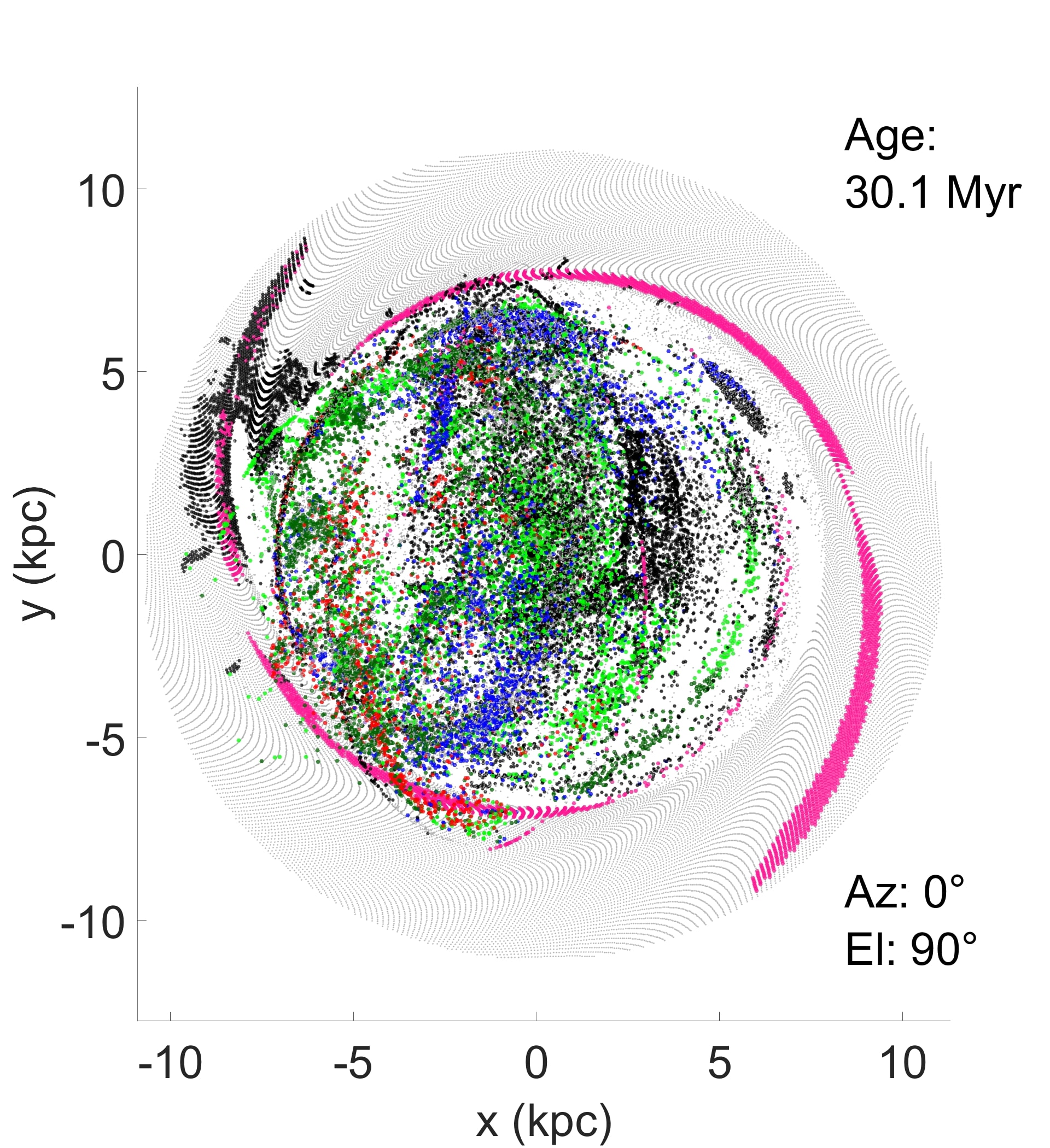}
        }\\ %  ------- End of the first row ----------------------%
                \subfigure[]{%
          \includegraphics[width=0.45\textwidth]{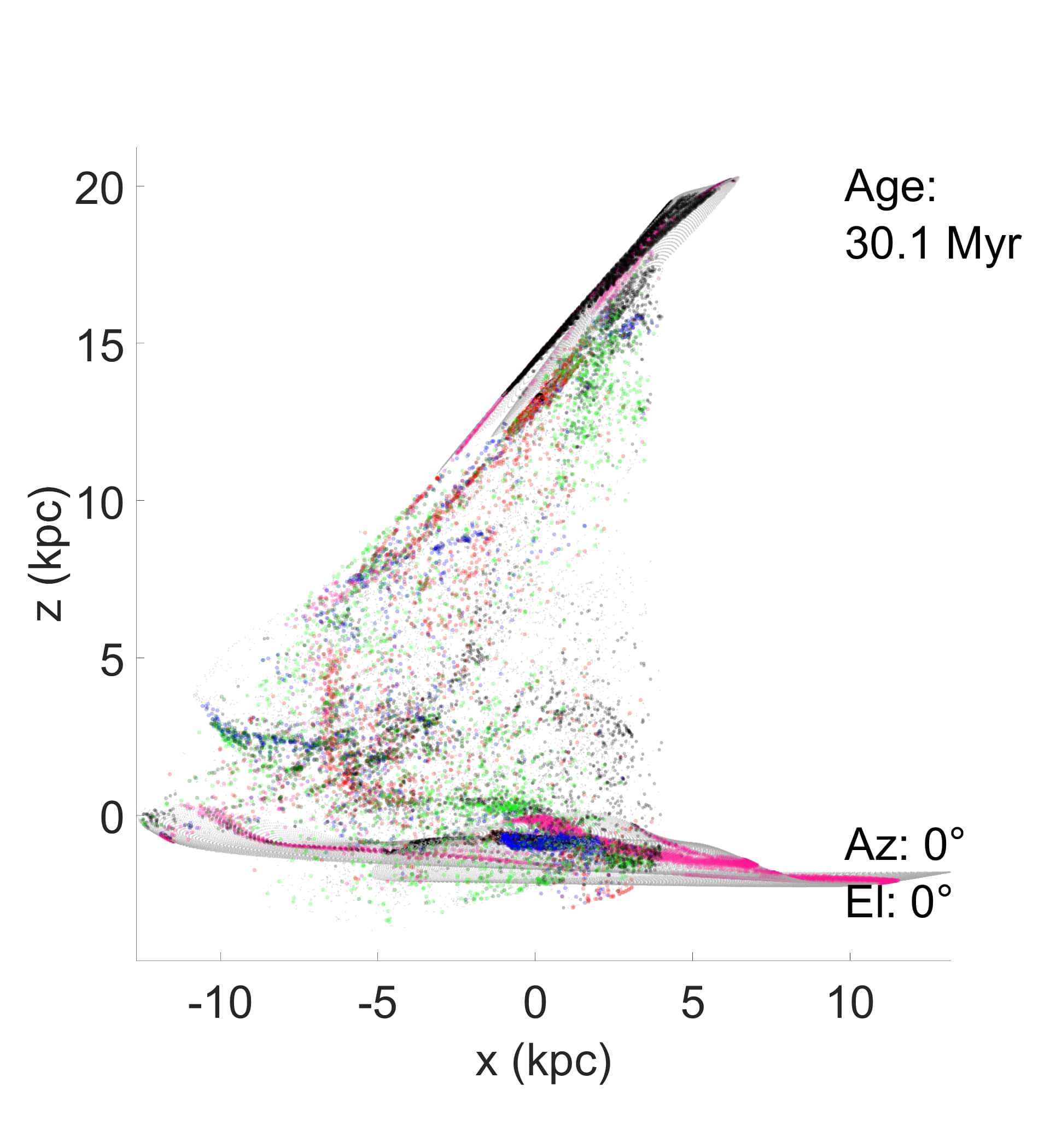}
        }
                \subfigure[]{%
          \includegraphics[width=0.45\textwidth]{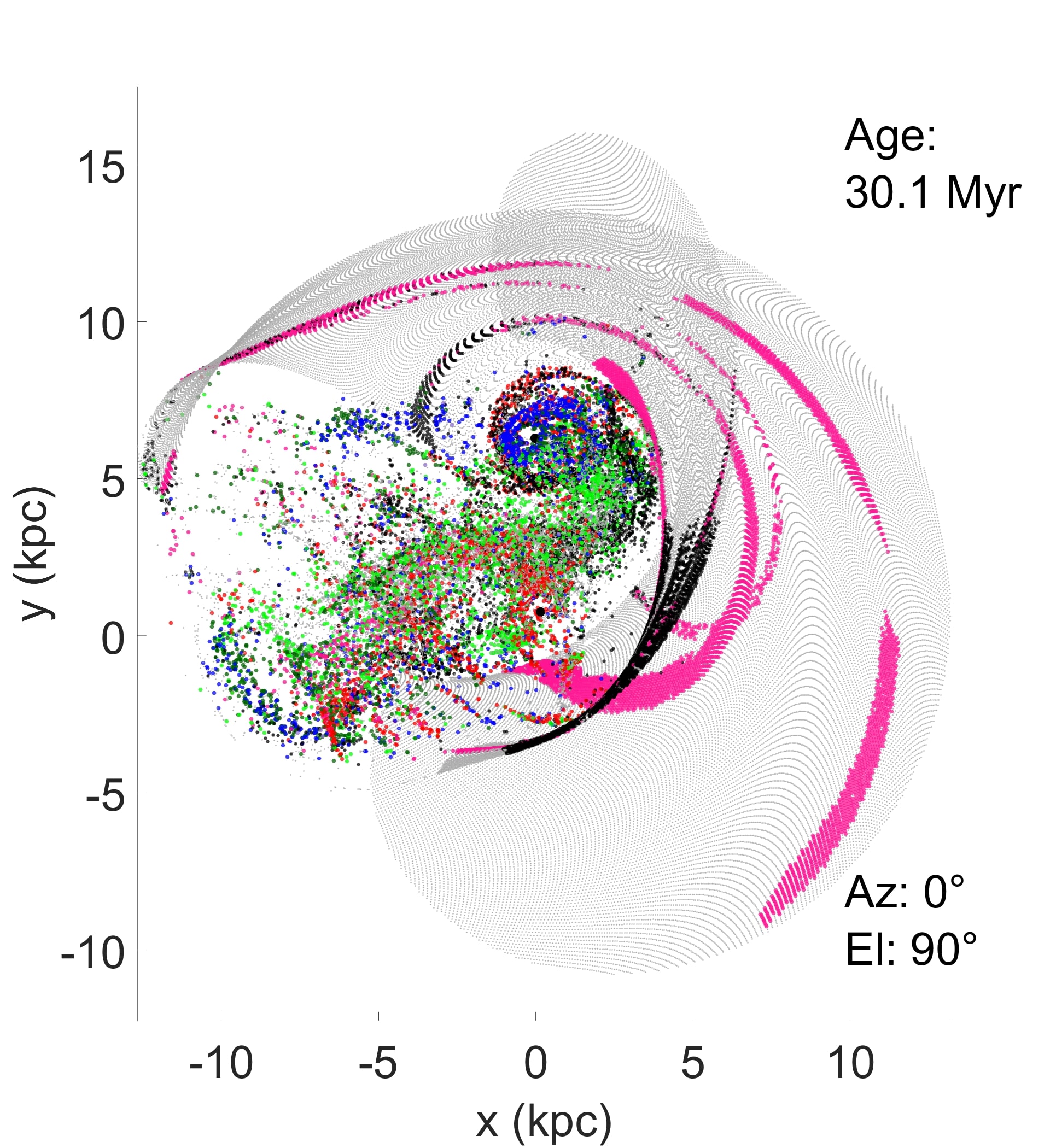}
        }\\ %  ------- End of the second row ----------------------%
\caption{Collisions with relative disk inclinations of $45\si{\degree}$.  Panels (a) and (b) are 500 pc offset collisions.  Panels (c) and (d) are 10 kpc offset collisions.  Color indicates when clouds finish collapse, and is explained in \cref{fig:sf0tilt1}.} %
\label{fig:sf45tilt1}
\end{figure*}

\begin{figure*}
    \centering
                \subfigure[]{%
          \includegraphics[width=0.45\textwidth]{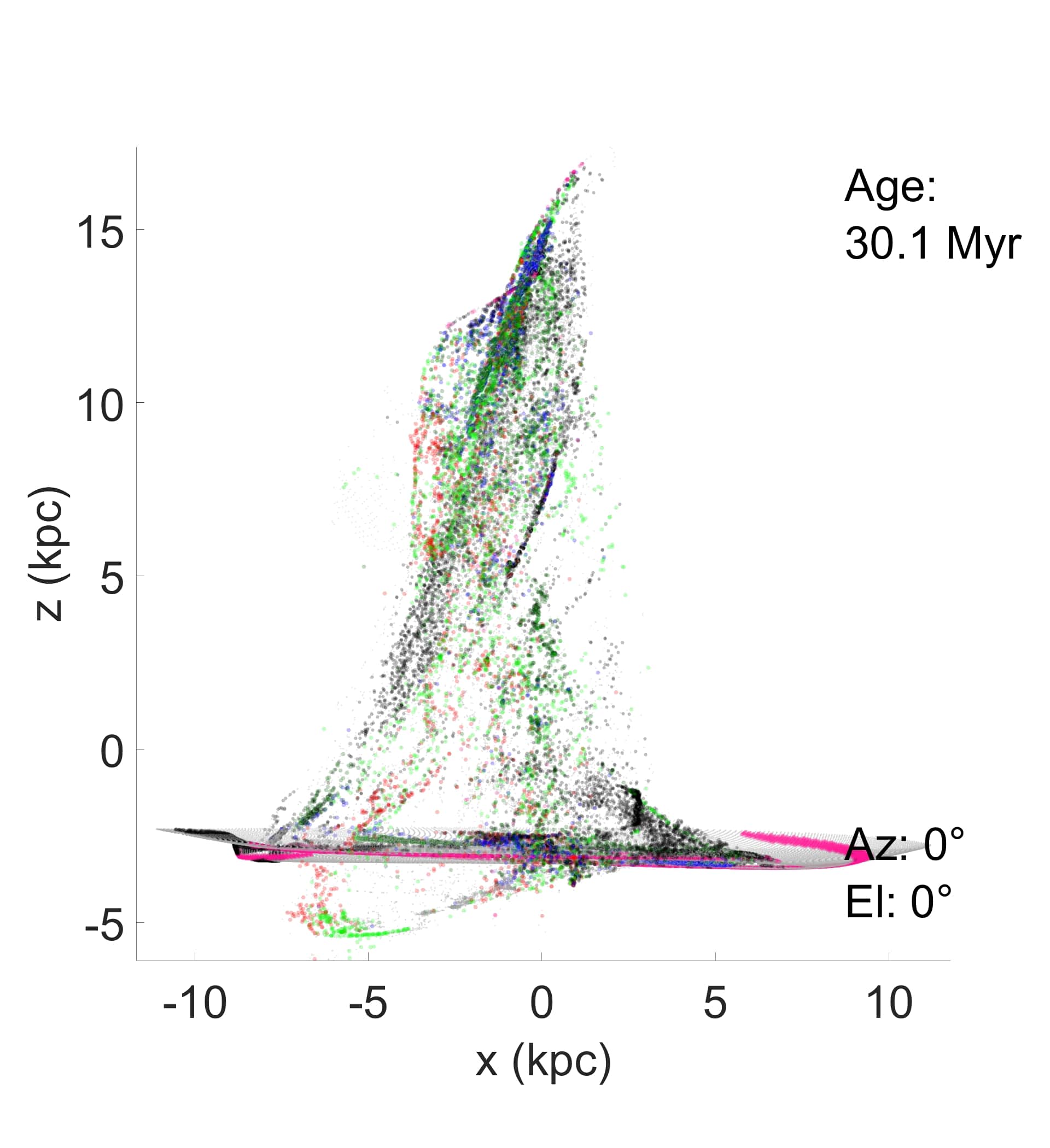}
        }
                \subfigure[]{%
          \includegraphics[width=0.45\textwidth]{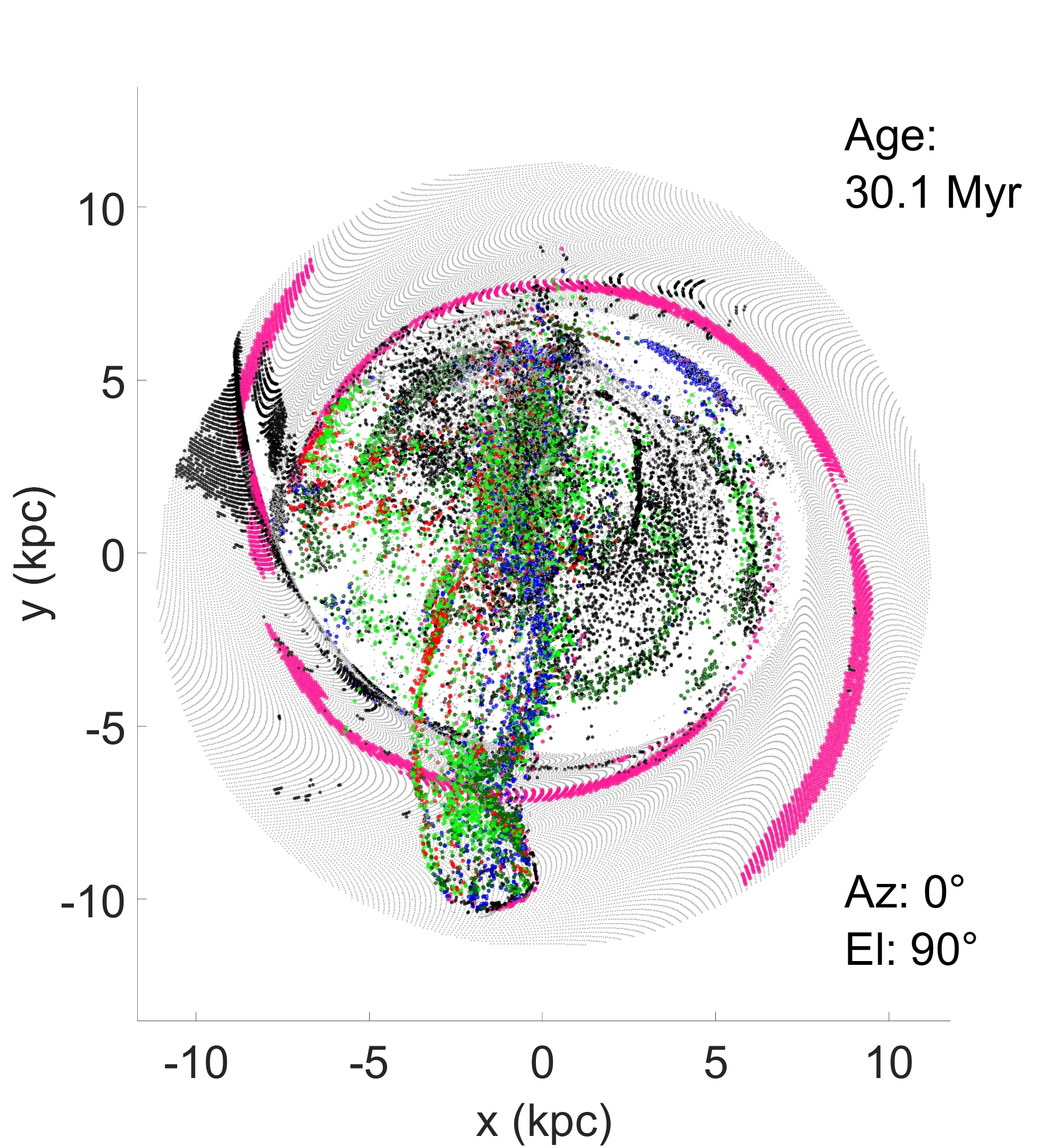}
        }\\ %  ------- End of the first row ----------------------%
                \subfigure[]{%
          \includegraphics[width=0.45\textwidth]{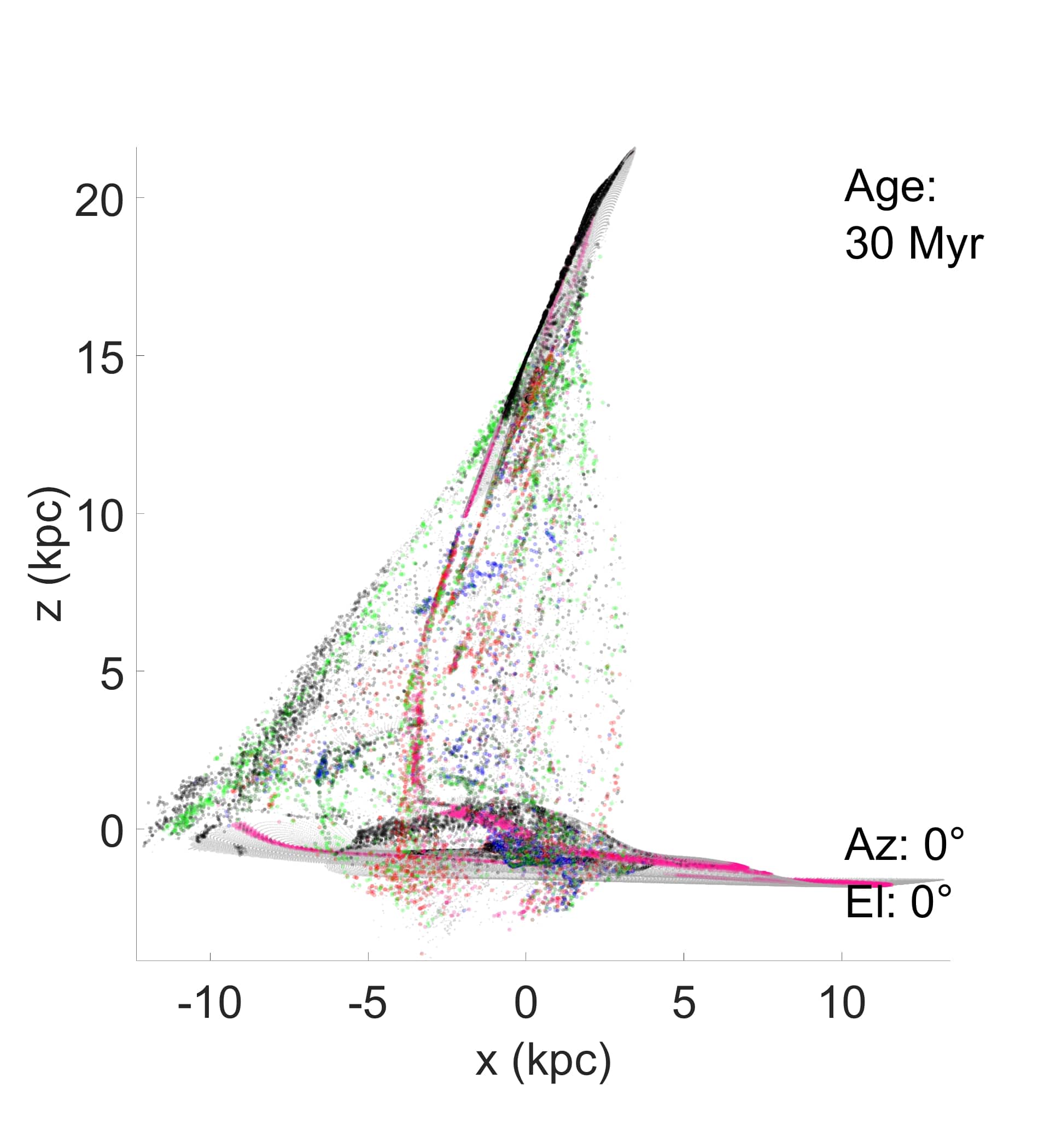}
        }
                \subfigure[]{%
          \includegraphics[width=0.45\textwidth]{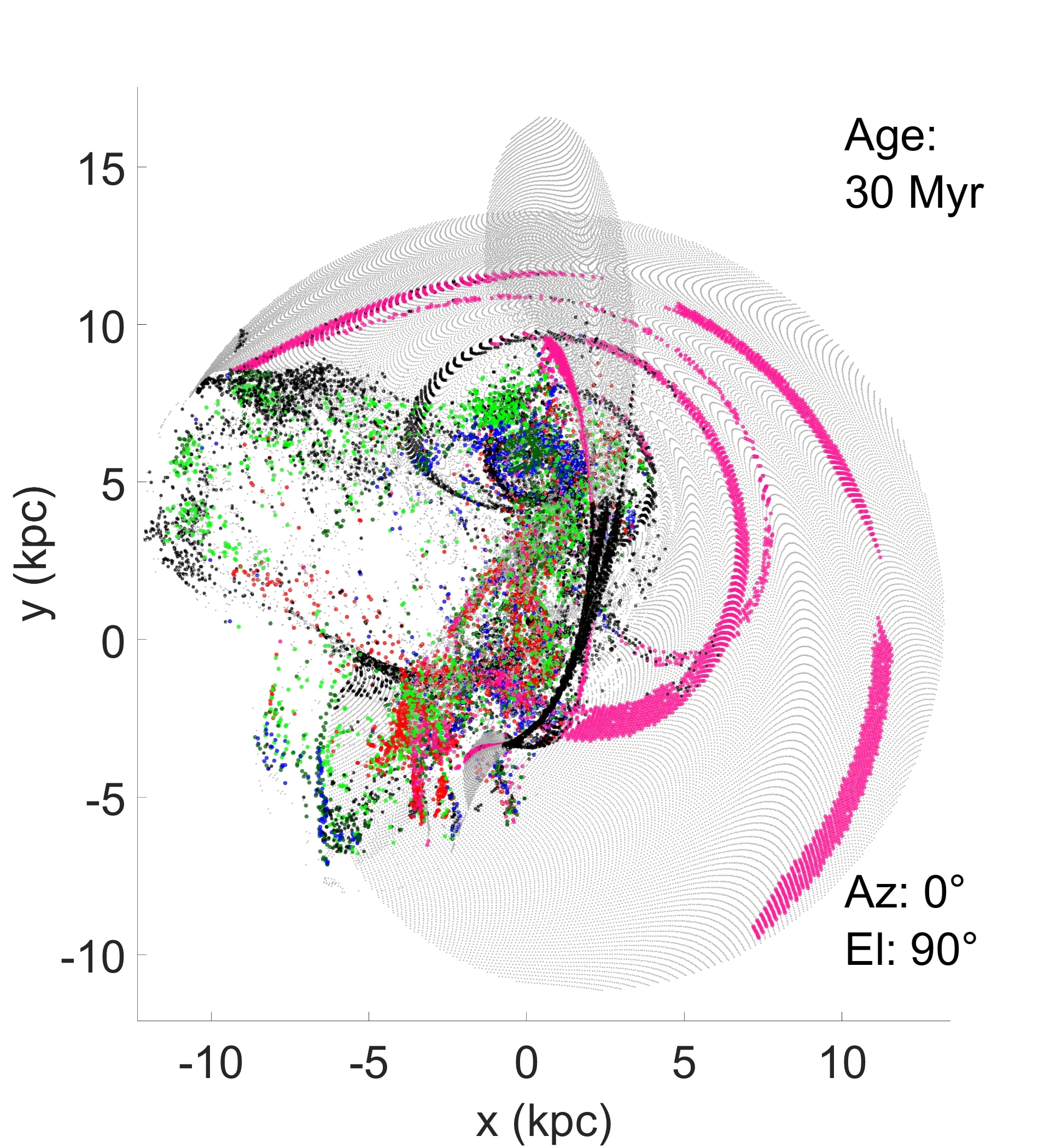}
        }\\ %  ------- End of the second row ----------------------%
\caption{Collisions with relative disk inclinations of $65\si{\degree}$.  Panels (a) and (b) are 500 pc offset collisions.  Panels (c) and (d) are 10 kpc offset collisions.  Color indicates when clouds finish collapse, and is explained in \cref{fig:sf0tilt1}.} %
\label{fig:sf65tilt1}
\end{figure*}

\begin{figure*}
    \centering
                \subfigure[]{%
          \includegraphics[width=0.45\textwidth]{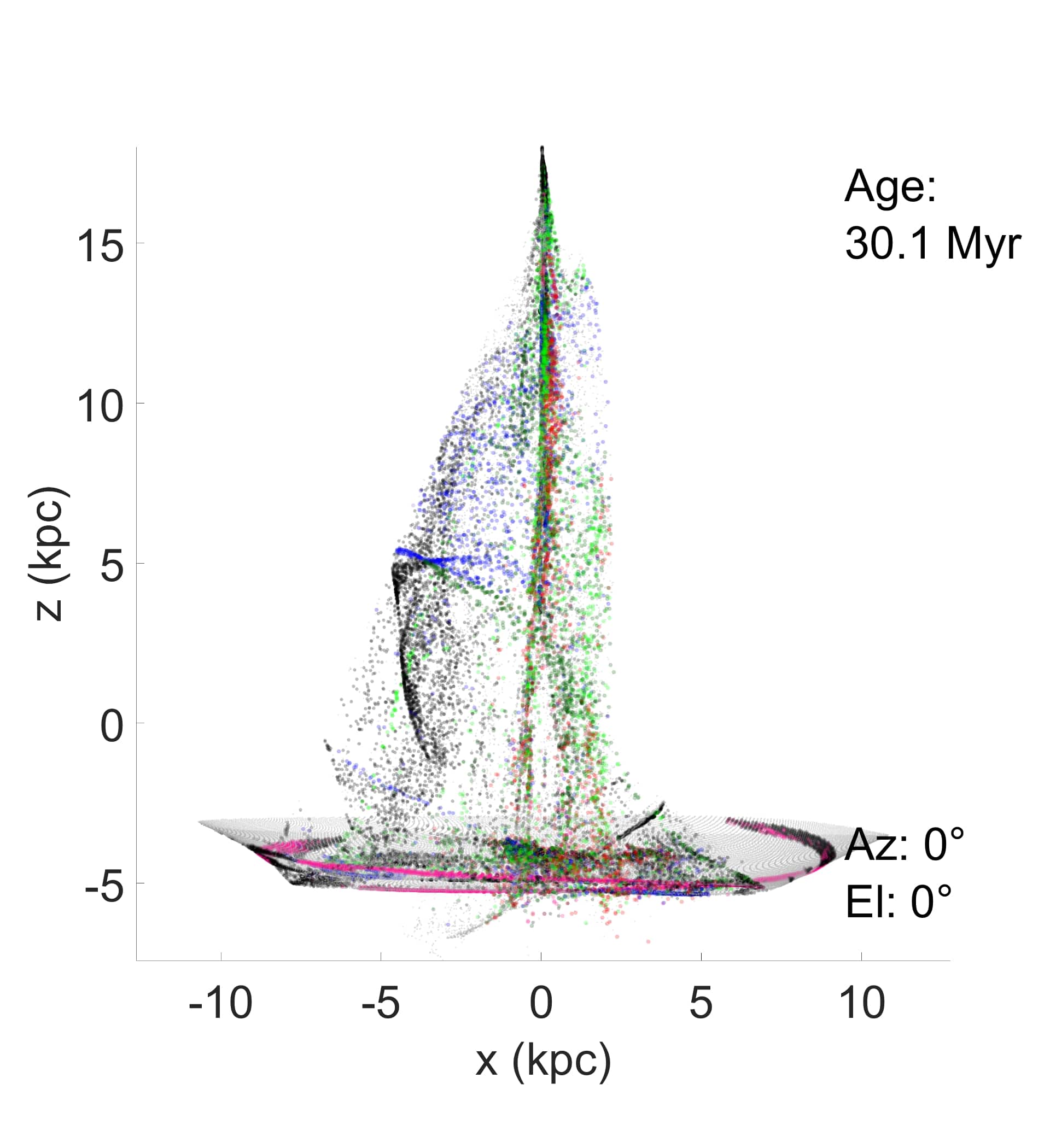}
        }
                \subfigure[]{%
          \includegraphics[width=0.45\textwidth]{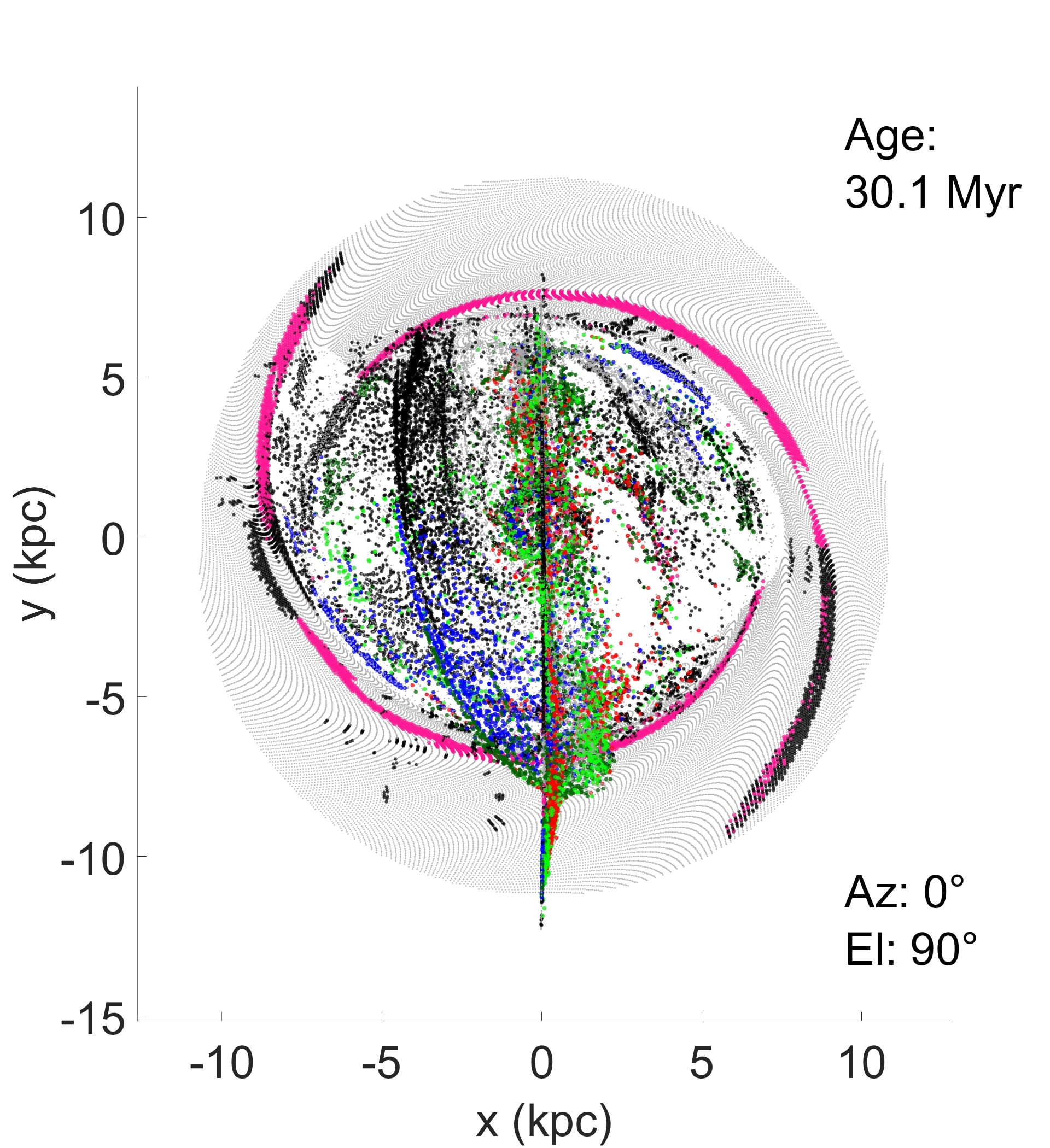}
        }\\ %  ------- End of the first row ----------------------%
                \subfigure[]{%
          \includegraphics[width=0.45\textwidth]{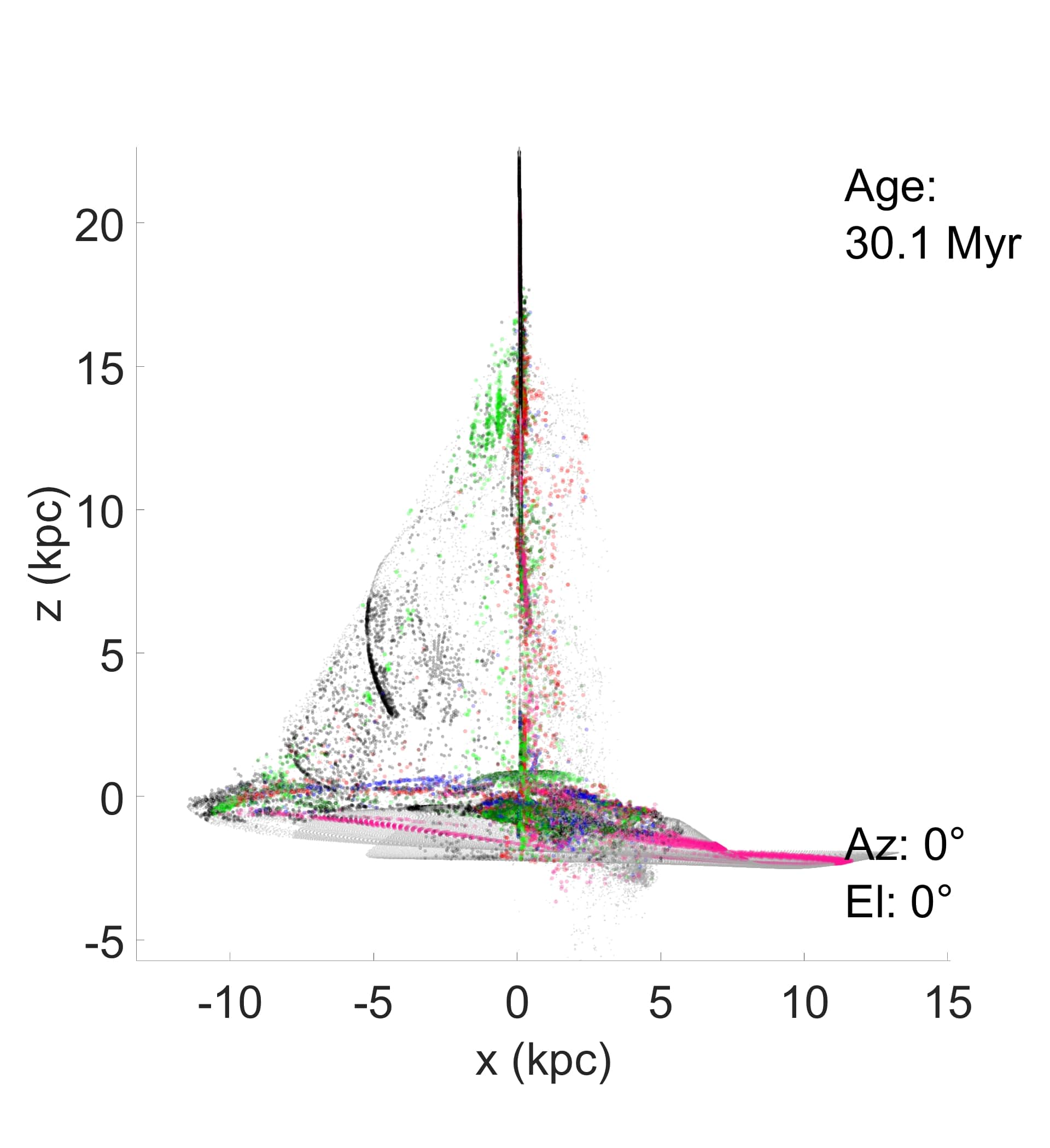}
        }
                \subfigure[]{%
          \includegraphics[width=0.45\textwidth]{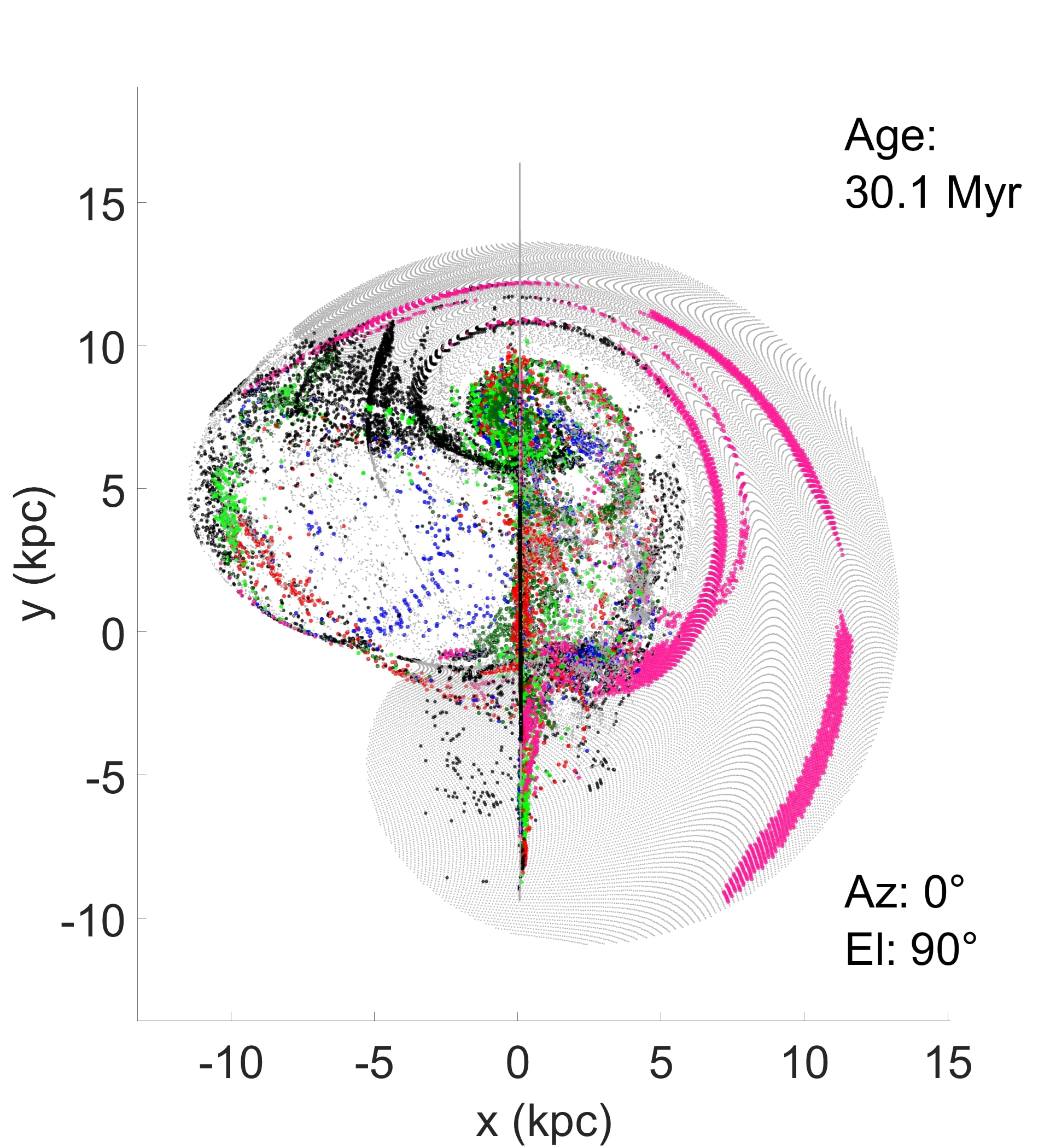}
        }\\ %  ------- End of the second row ----------------------%
\caption{Collisions with relative disk inclinations of $90\si{\degree}$.  Panels (a) and (b) are 500 pc offset collisions.  Panels (c) and (d) are 10 kpc offset collisions.  Color indicates when clouds finish collapse, and is explained in \cref{fig:sf0tilt1}.} %
\label{fig:sf90tilt1}
\end{figure*}

\indent \Cref{fig:sf0tilt1,fig:sf20tilt1,fig:sf45tilt1,fig:sf65tilt1,fig:sf90tilt1} show snapshots with color-coded cloud collapse information, in addition to the spatial distribution of clouds at the given times.  Pink dots represent clouds that collapsed at times $>$10 Myr before disk-disk collision, red clouds 10 Myr or less before the disk-disk collision.  Bright green dots collapsed at 0-10 Myr after closest approach, dark green dots at 10-20 Myr, blue dots at 20-30 Myr, purple at 30-120 Myr (end of run).  Black dots indicate gas clouds that are unstable but have not completed a freefall time by the end of the run.  Light gray dots (barely visible) are gas clouds that never become gravitationally unstable. The time since the disk-disk collision is listed in the top right of each panel of these figures.  The azimuth and elevation of the point of view is given in the bottom right of each panel of all figures. These figures show that many of the (assumed) 100 pc diameter gas clouds involved in the direct disk-disk collisions eventually reach the Jeans limit for gravitational collapse.  This implies splash bridges, regardless of impact parameter, should trigger star formation.  However, the impact parameters affect the detailed star formation histories.  

\indent Low-inclination collisions are shown in \cref{fig:sf0tilt1,fig:sf20tilt1} at an age of 30 Myr.  This age is chosen since it is the upper age estimate for the time since the disk-disk collision of the Taffy Galaxy system.  The low-inclination collisions do not produce many gas cloud collapses in the bridge before 20 Myr.  But, as shown with blue dots, there are a number of massive clumps that complete gravitational collapse between 20 and 30 Myr in the bridge centers.  All of the blue clouds are ready to start forming stars across the bridge, triggering a starburst.  There is a sizeable amount of gas that does not finish collapse by 30 Myr, though there are not enough clouds in this state to explain the absence of star formation in the Taffy bridge.  We propose that the Taffy bridge exists near the end of a dark period between star bursts.  

\indent There is an approximately 3 kpc region of gas colored green in panel a) of \cref{fig:sf20tilt1} just below the upper galaxy G2.  This region of gas finishes collapsing 10 Myr after the collision, and is in a similar state as the active dense region of \hh{} seen in the Taffy bridge.  As a general trend, low-inclination collisions produce the most delayed cloud collapses of all splash bridges.  The longer delay is likely due to the fact that the strongest initial shocks are produced in counter-rotating low-inclination collisions.  Shock waves in the low-inclination collisions are able to drive much more of the gas up to million degree temperatures, yielding long cooling times.

\indent An interesting pattern of cloud collapse is seen in panel b) of \cref{fig:sf0tilt1}.  There are several rings of particles with different cloud collapse times.  From the center of the galaxy outward there is a $>$40 Myr collapse time ring, 30 Myr clumpy ring, 10 Myr green ring and finally another $>$40 Myr old ring.  These rings appear to be contained in a single disk, however, the edge-on view of panel a) of this figure tells a different story.  One of the $>$40 Myr old rings along with the 30 Myr ring are in fact in the center of the bridge.  The outer $>$40 Myr ring is a ring wave in the companion G2.  The 10 Myr ring has clouds in the disk of galaxy G1 as well as just below the disk of G2.   In \cref{fig:sf20tilt1} there are still some rings of cloud collapse but the 20\si{\degree} inclination causes a several Myr delay in cloud collapse across the gas disk.  

\indent Collisions with higher relative disk inclinations of 45\si{\degree}, 65\si{\degree} and 90\si{\degree} are shown in \cref{fig:sf45tilt1,fig:sf65tilt1,fig:sf90tilt1}.  At these inclinations there is a great spread in gas cloud collapse times.  Many stars are formed before the disk-disk collision.  This is because the inclination leads to one edge of the gas disk of G1 intersecting G2 several million years before the closest approach of centers.  Turbulence in the splash bridge keeps a low but steady rate of gas cloud collapse going for up to 70 Myr after the collision.  

\indent The 45$\si{\degree}$ angle disk-disk collision differs visually from the other collisions.  The bridge in this case does not develop a prominent central bridge disk and does not form a ring of gas in the bridge.  The bridge also does not produce the streams of gas near G2 as seen in higher inclination collisions.  The gas remaining in the disks of G1 and G2 does, however, experience ring waves after the disk-disk collision, which is similar to the low-inclination collisions.

\begin{figure*}
    \centering
                \subfigure[]{%
          \includegraphics[width=0.45\textwidth]{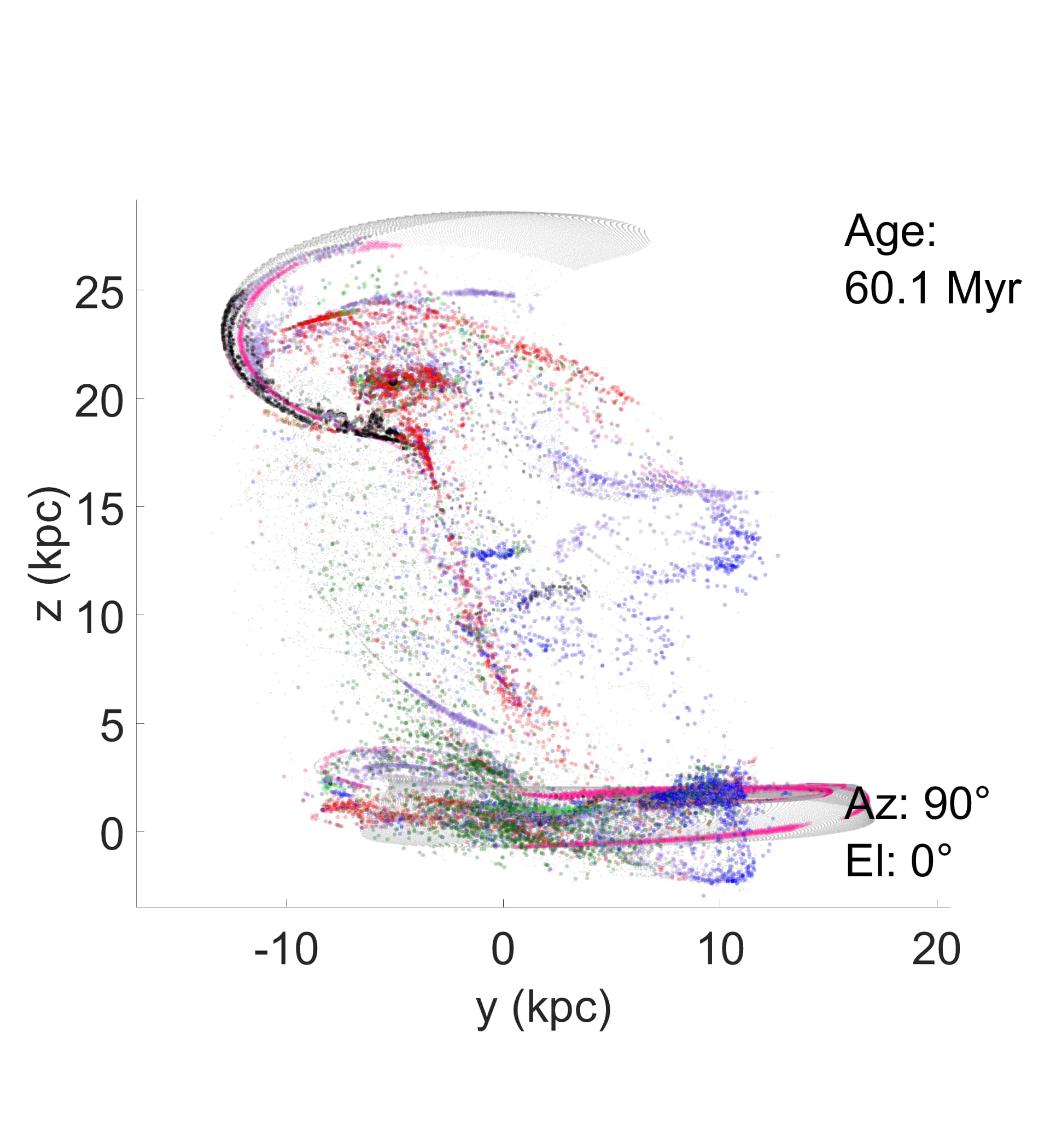}
        }
                \subfigure[]{%
          \includegraphics[width=0.45\textwidth]{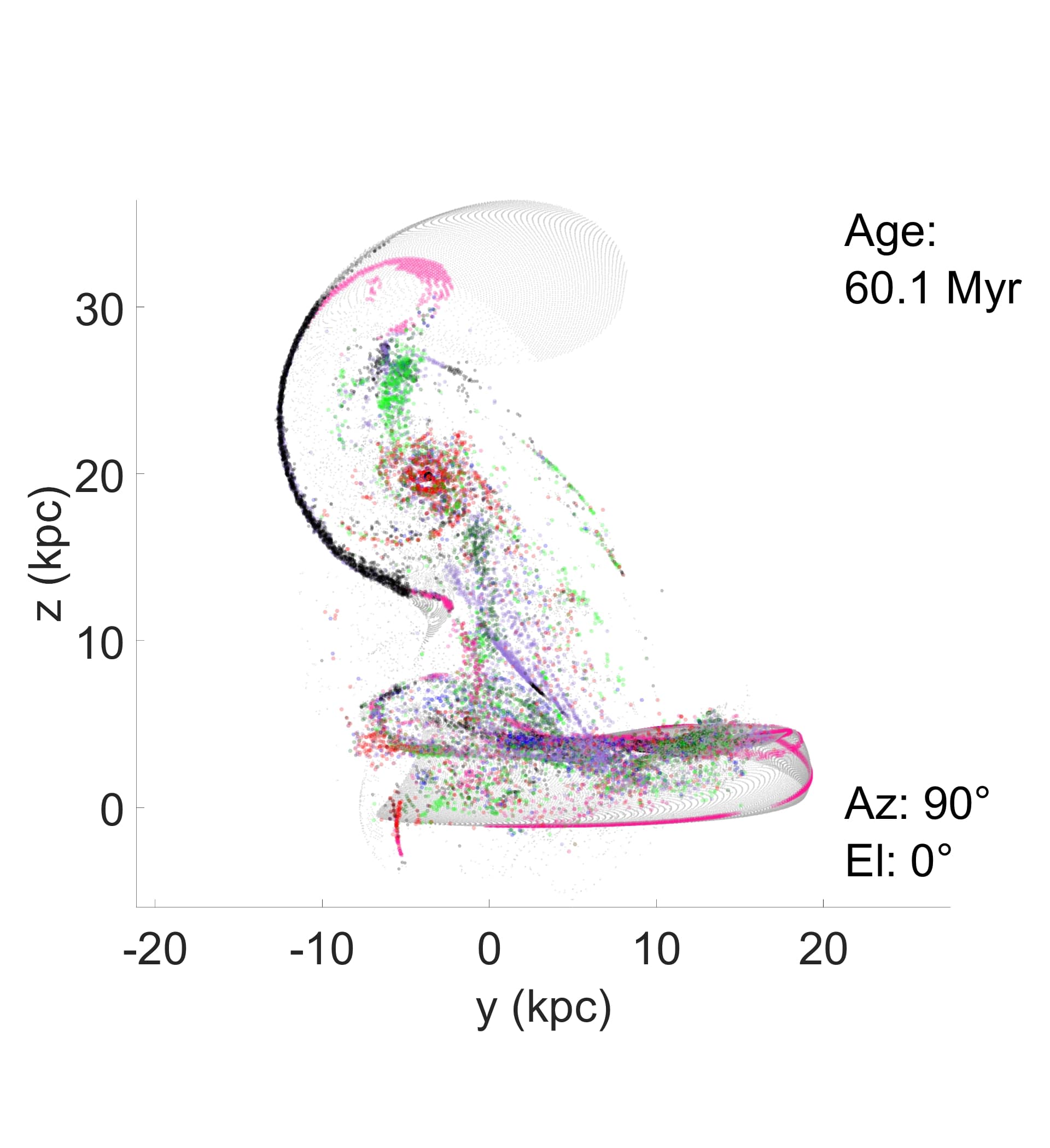}
        }\\ %  ------- End of the first row ----------------------%
                \subfigure[]{%
          \includegraphics[width=0.45\textwidth]{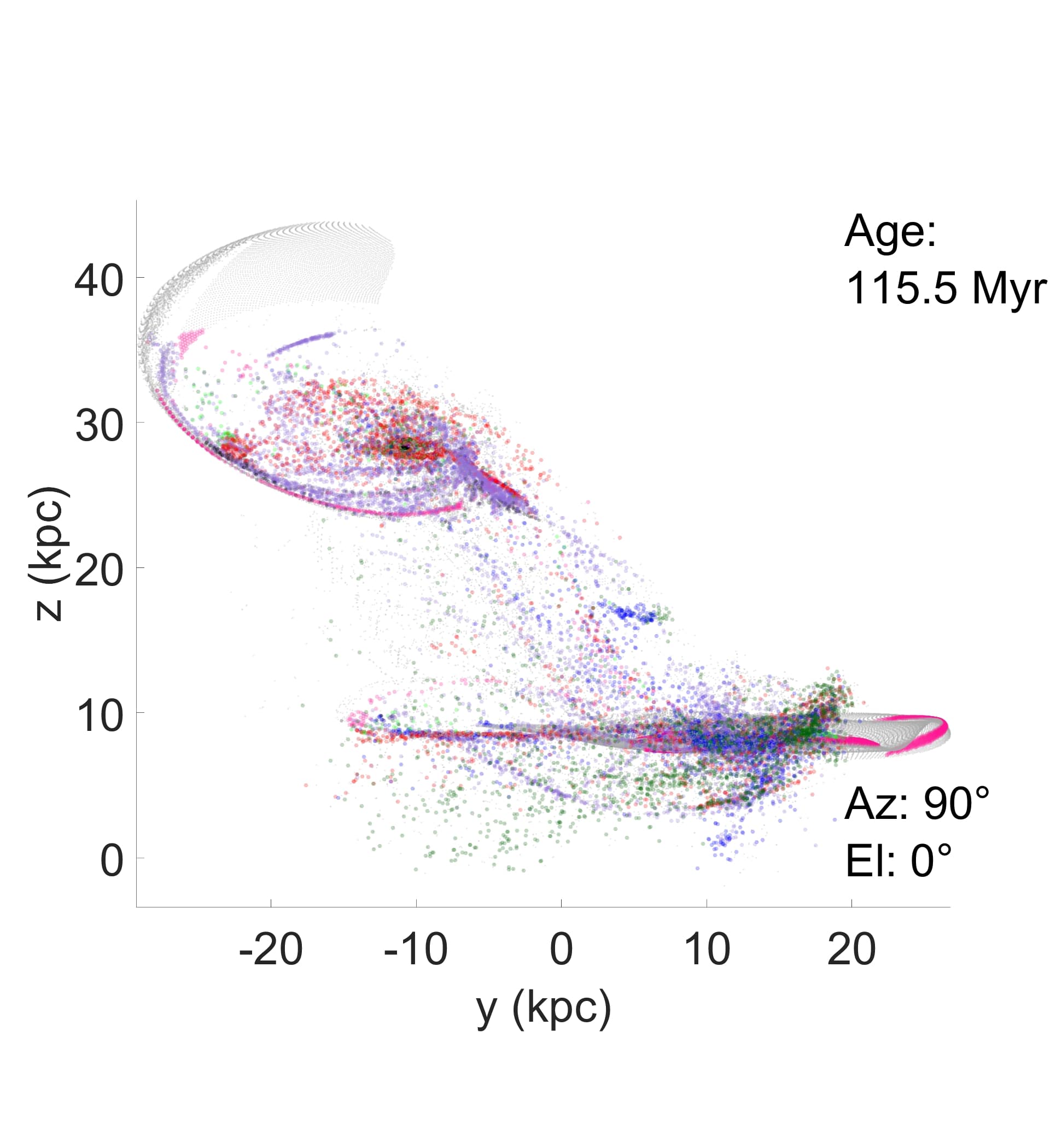}
        }
                \subfigure[]{%
          \includegraphics[width=0.45\textwidth]{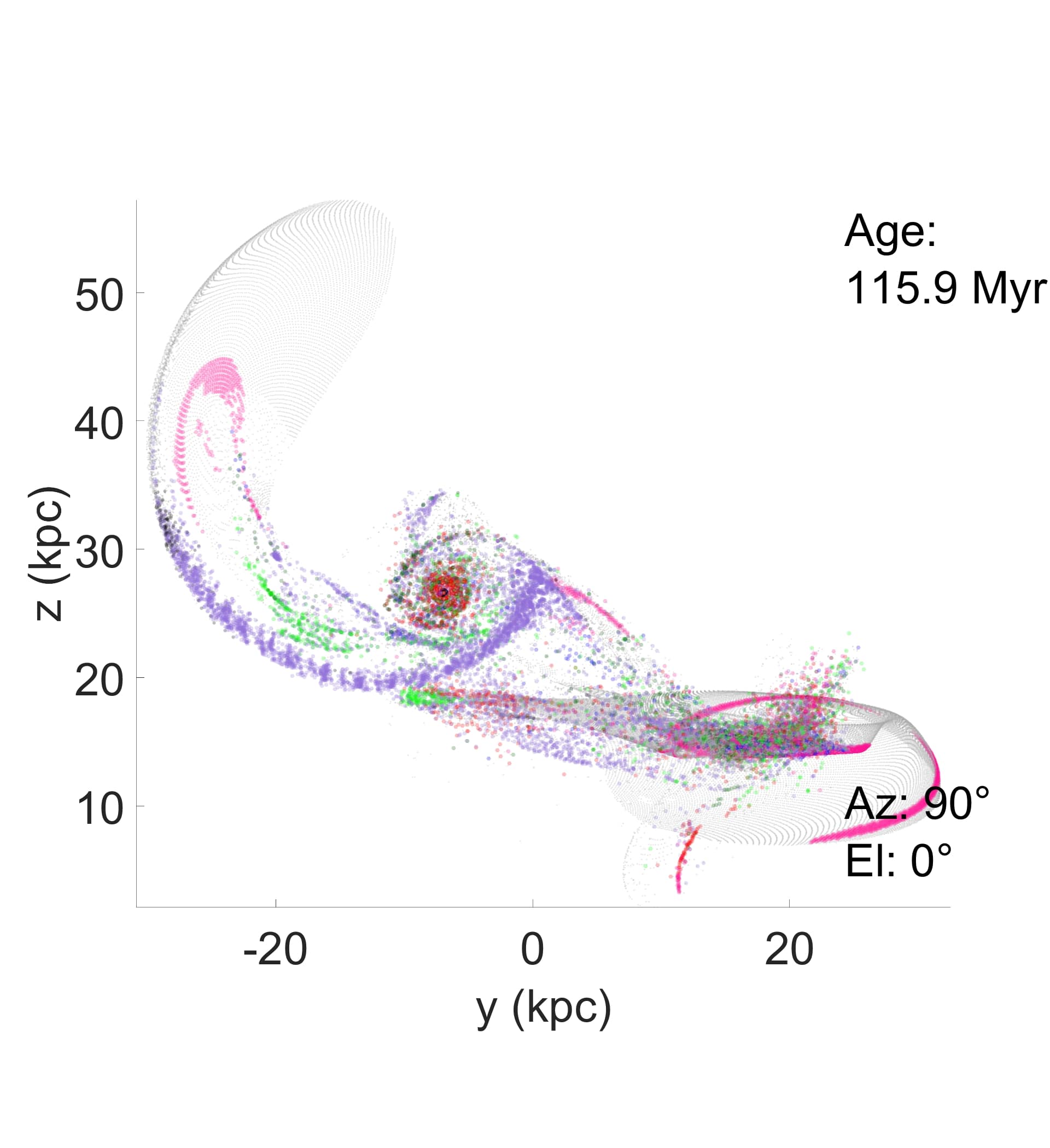}
        }\\ %  ------- End of the second row ----------------------%
\caption{10 kpc offset runs.  Panel (a) and (c) are 20\si{\degree} inclination collisions and panel (b) and (d) are 90\si{\degree} inclination collisions.  Color indicates when clouds finish collapse, and is explained in \cref{fig:sf0tilt1}.} %
\label{fig:sflatestage}
\end{figure*}

\Cref{fig:sflatestage} shows both low and high-inclination collisions, with 10 kpc impact offset, at two relatively late times.  This shows the re-accretion of gas back onto G1 and G2.  Previously this gas was only considered to be reheated when it fell back onto the two galaxies, but in fact many of the gas clouds become gravitationally unstable and have completed a freefall time before the re-accretion.  Star clusters of various ages will also fall back with the leftover gas.  In the 60 Myr panels (a) and (b) a massive, tens of kiloparsecs long filament, shown by black dots, still has not produced stars.  This filament of gas clouds ultimately collapses nearly 80 Myr after the disk-disk collision.  The unstable clouds in the bridge of gas shown in \Cref{fig:sflatestage} is similar to the observed star forming regions in Arp 194.  

\begin{figure*}
     \centering
                \subfigure[]{%
          \includegraphics[width=0.45\textwidth]{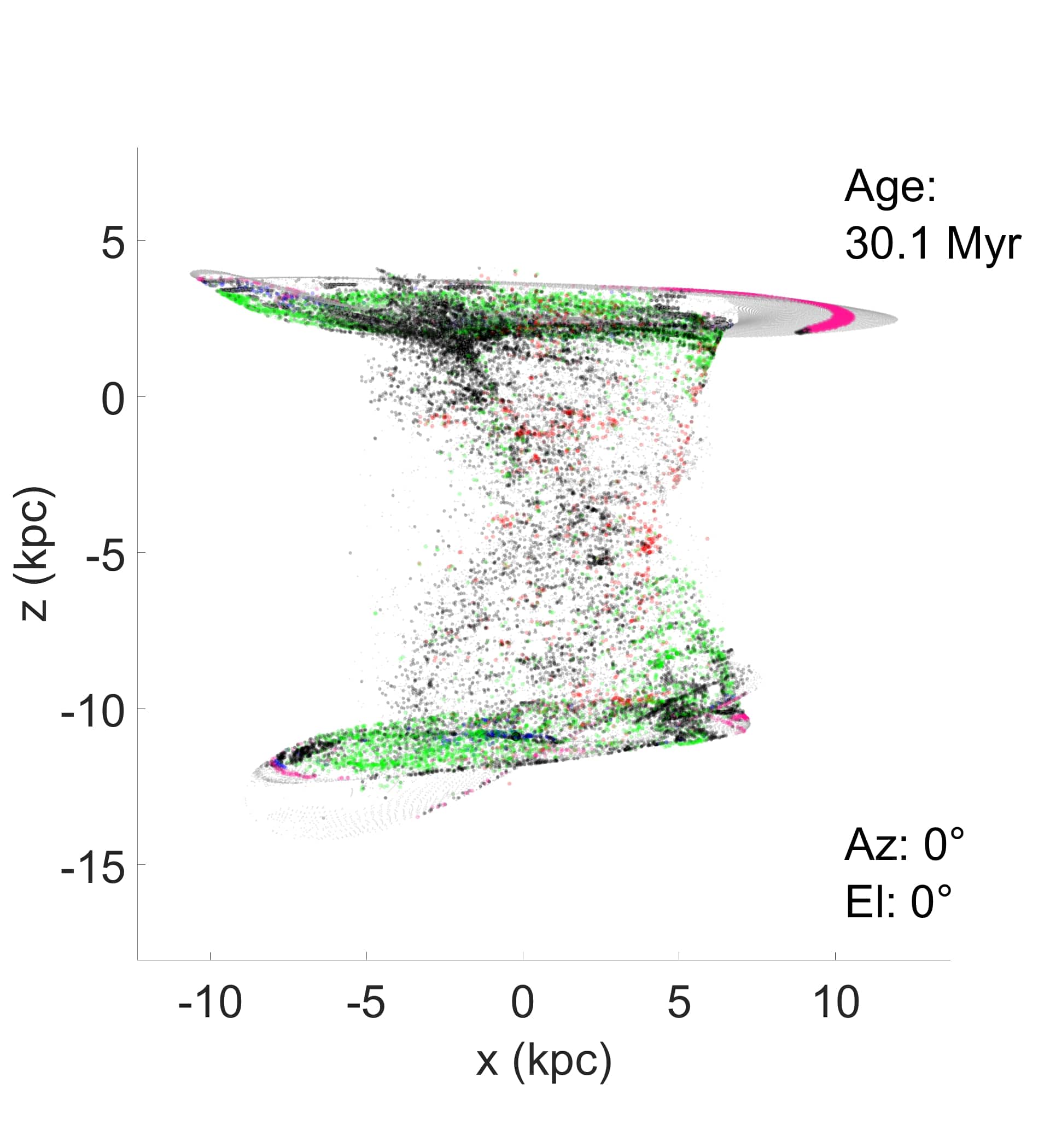}
        }
                \subfigure[]{%
          \includegraphics[width=0.45\textwidth]{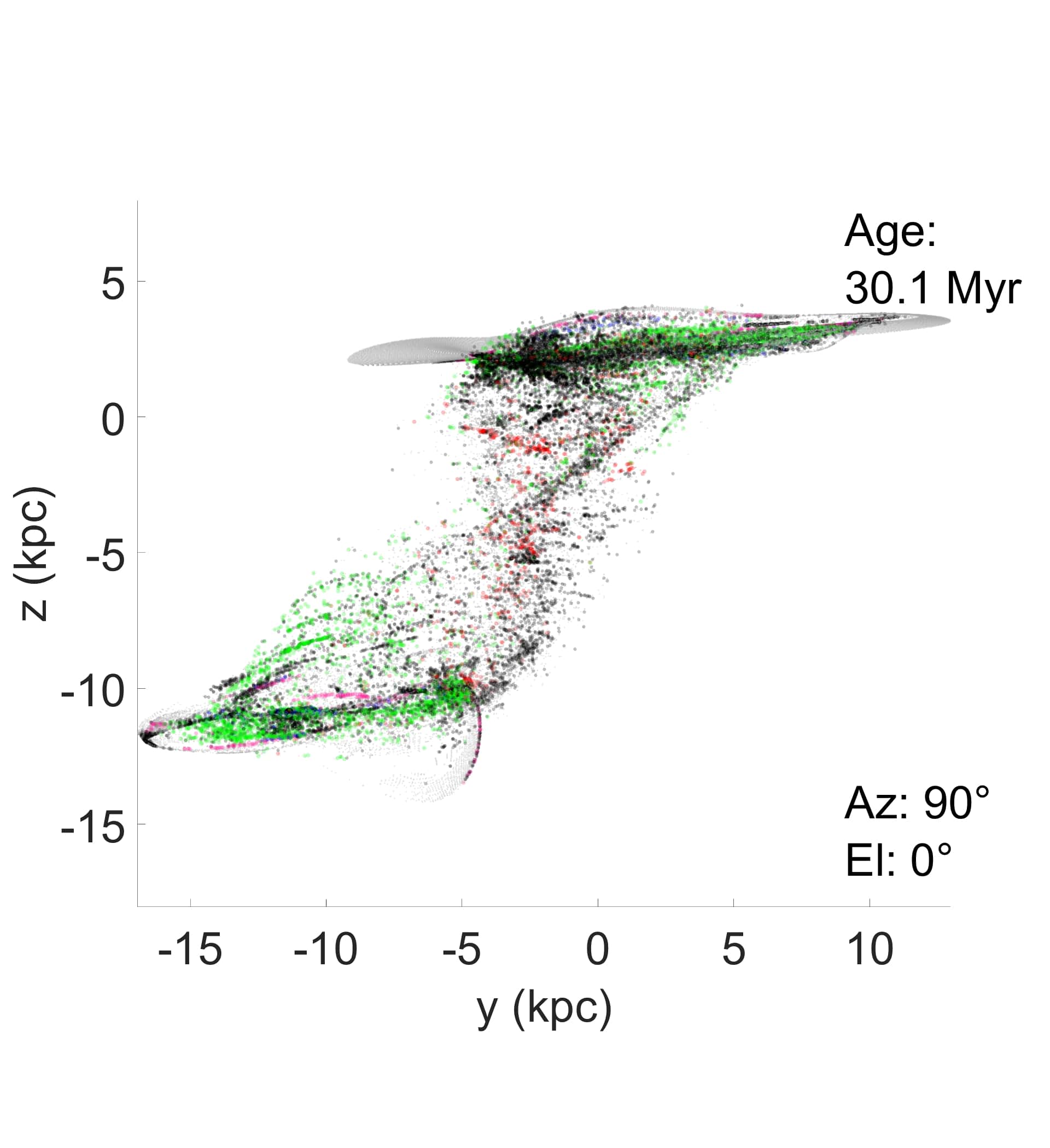}
        }\\ %  ------- End of the first row ----------------------%
        \subfigure[]{%
          \includegraphics[width=0.45\textwidth]{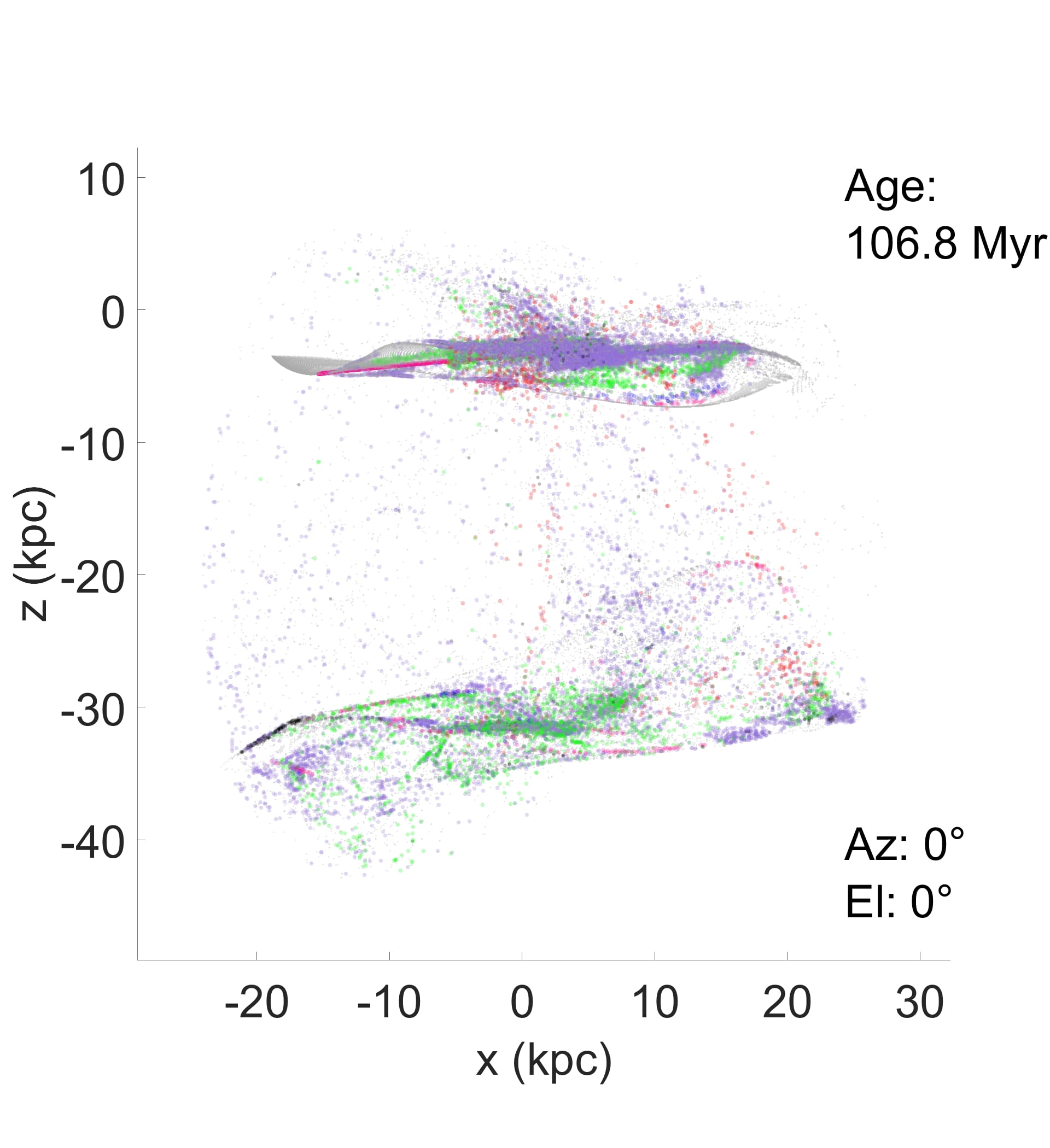}
        }
                \subfigure[]{%
          \includegraphics[width=0.45\textwidth]{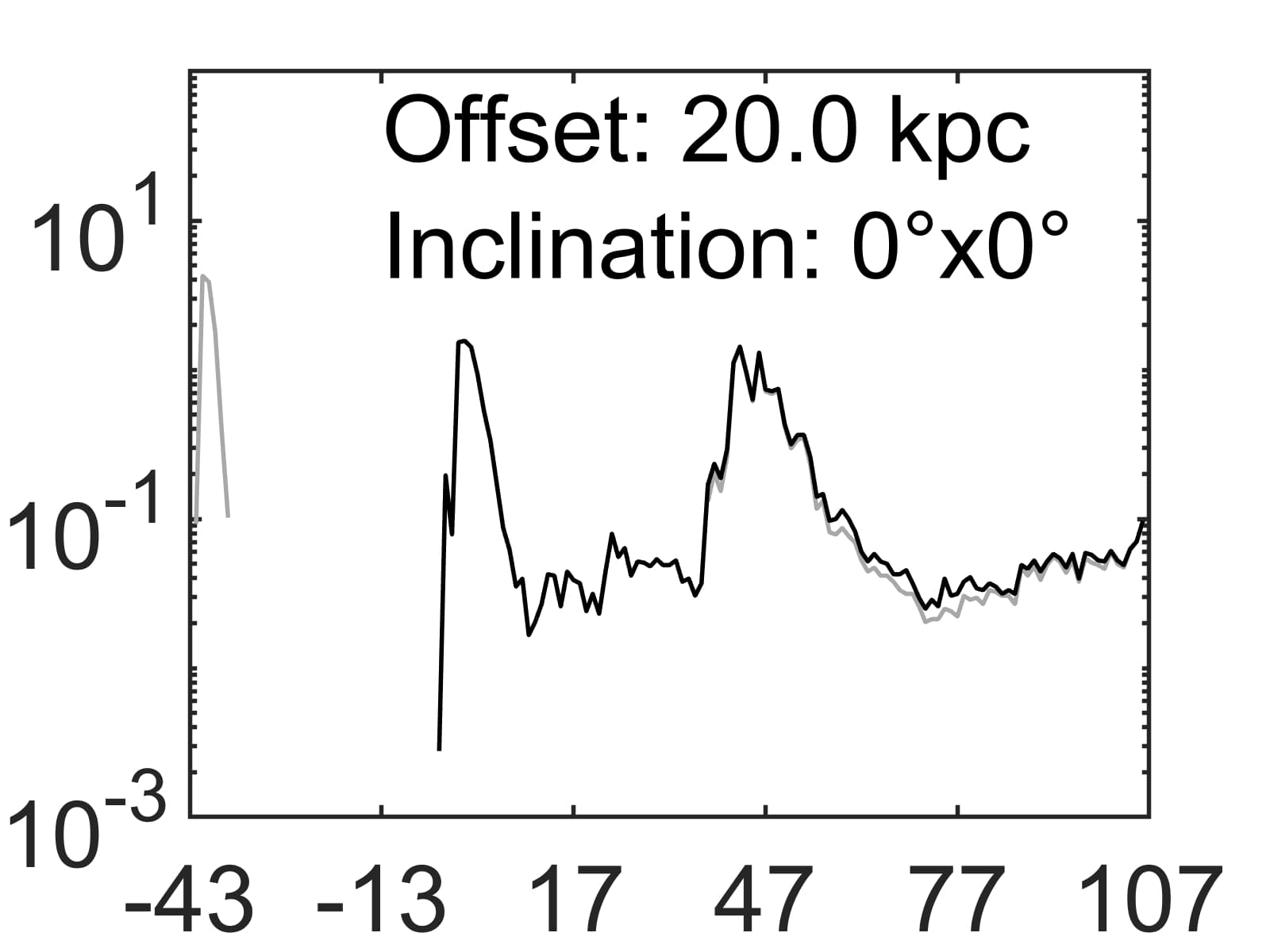}
        }\\
\caption{Collision was initialized with G2 offset 20 kpc on the y-axis and 20 kpc up the z-axis from G1 and the disks are not initially inclined to one another.  The initial velocity of G2 was set to 350 km/s in the $-y$ direction and 350 km/s in the $-z$ direction.  The two galaxies accelerate from their initial positions, reaching a 1000 \kms{} impact velocity with one another.  G1 is above G2 in these frames, in contrast to the other runs.  Color in panels (a), (b), and (c), indicates when clouds finish collapse, and is explained in \cref{fig:sf0tilt1}.  Panel (d) is the fractional cloud collapse rate versus time, to be compared with \cref{fig:sfrate1}.} %
\label{fig:sf0tiltglancing}
\end{figure*}

\indent \Cref{fig:sf0tiltglancing} shows a 30 Myr old splash bridge formed from a glancing collision between two galaxies.  The collision velocity in this run was over 1000 \kms{}.  The increase in collision velocity results in a further delay before the gas clouds can collapse.  Recall that the low-inclination models whose disks collided with velocities ranging from 625 to 900 \kms{} have gas cloud collapse delayed for 20-30 Myr.  Collision velocities beyond 1000 \kms{} drive much of the gas into the thermal bremsstrahlung region of cooling.  The cooling timescales for \HI{} gas of density order \SI{e7} \dunit{} and T $>$ \SI{e7} grow extremely quickly as seen in \cref{fig:coolingtimetable20000}.  Panel d) shows the fractional gas cloud collapse rate over time for this collision.  The 50 Myr delay in gas cloud collapses is clearly seen as a secondary burst in the graph.  Panel c) which shows the collision at an age of 106.8 Myr a time when most of the collapsed gas clouds have accreted back onto one of the two galaxies in both cases.  There are a good number of gas clouds that collapse within the first 10 Myr after the collision, which are located in a 5 kpc sized region next to the disk of G2.  The region is visible in panel b) of \cref{fig:sf0tiltglancing}.  This sparsely populated clump of collapsed clouds is in a similar location to the clump observed by UGC 12915.  Yet this model bridge is still not a perfect fit to the Taffy bridge, since there are some collapsing gas clouds below G1 as well. These clouds collapsed 10 Myr before the disk-disk collision.  Many of the collapsed gas clouds are completely surrounded by other stable gas clouds, which could obscure diffuse star formation through the bridge.

\subsection{Distributions of young stars}

\begin{figure*}
     \begin{center}
                \subfigure[]{%
          \includegraphics[width=0.45\textwidth]{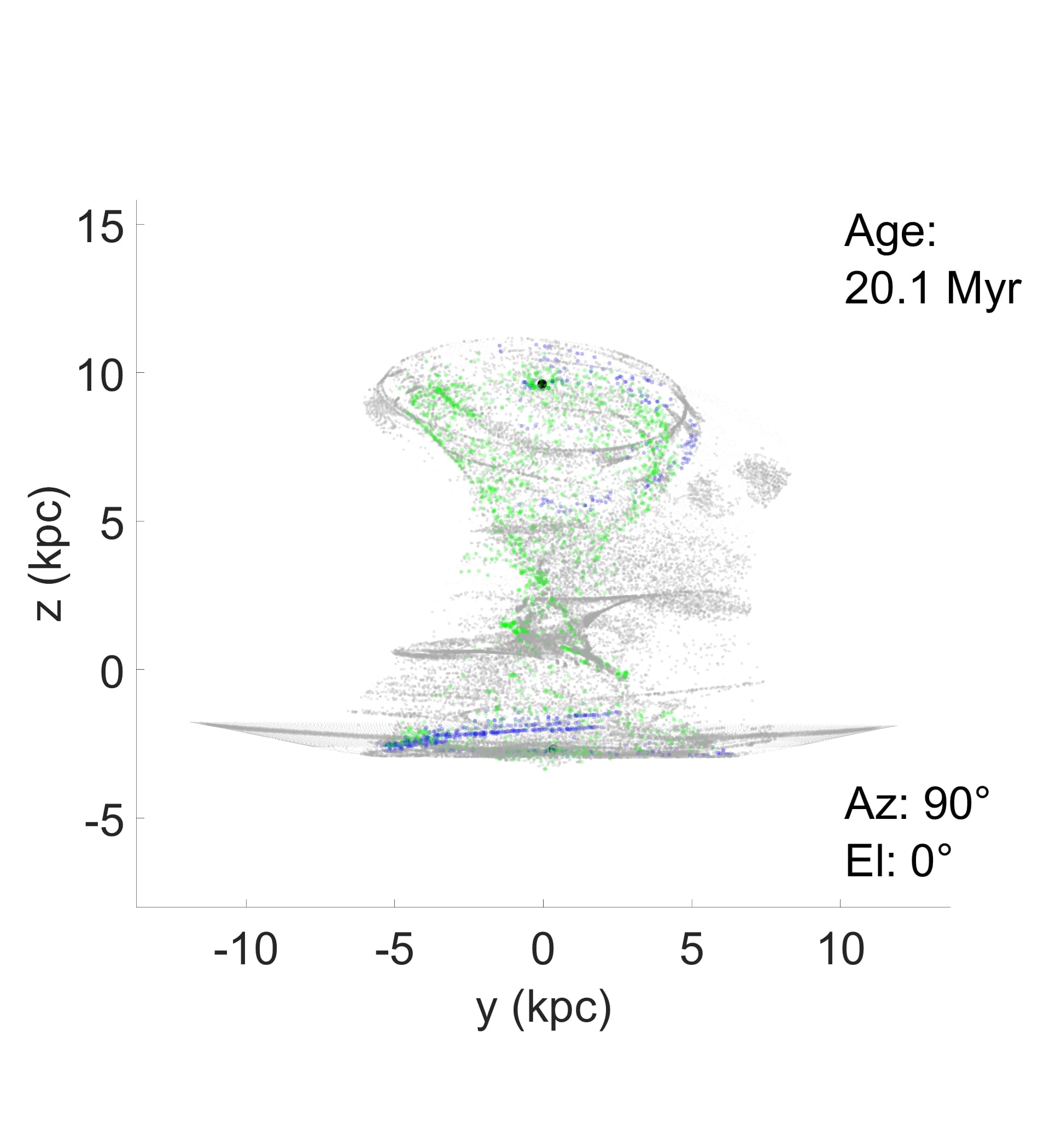}
        }
                \subfigure[]{%
          \includegraphics[width=0.45\textwidth]{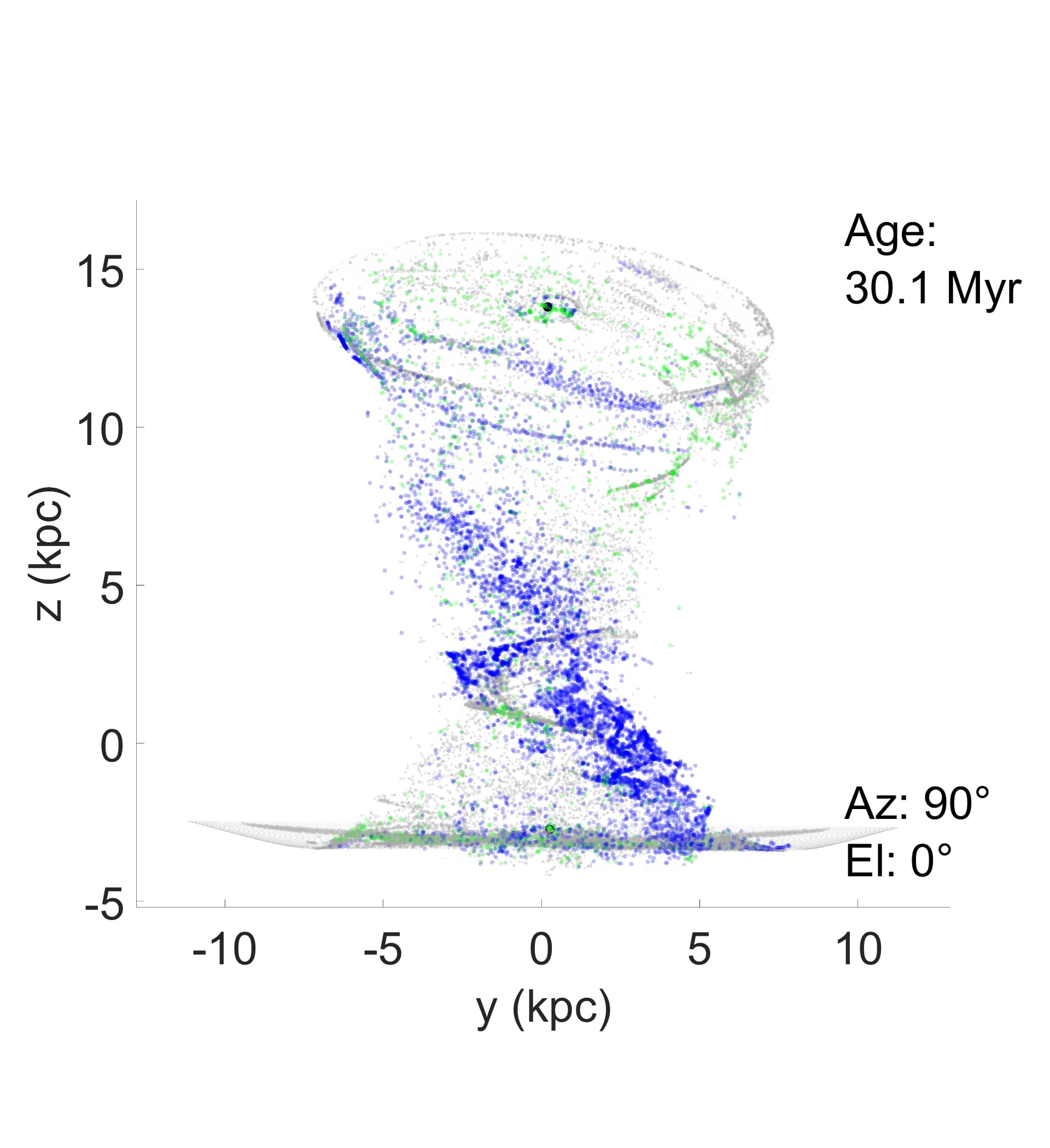}
        }\\ %  ------- End of the first row ----------------------%
        \subfigure[]{%
          \includegraphics[width=0.45\textwidth]{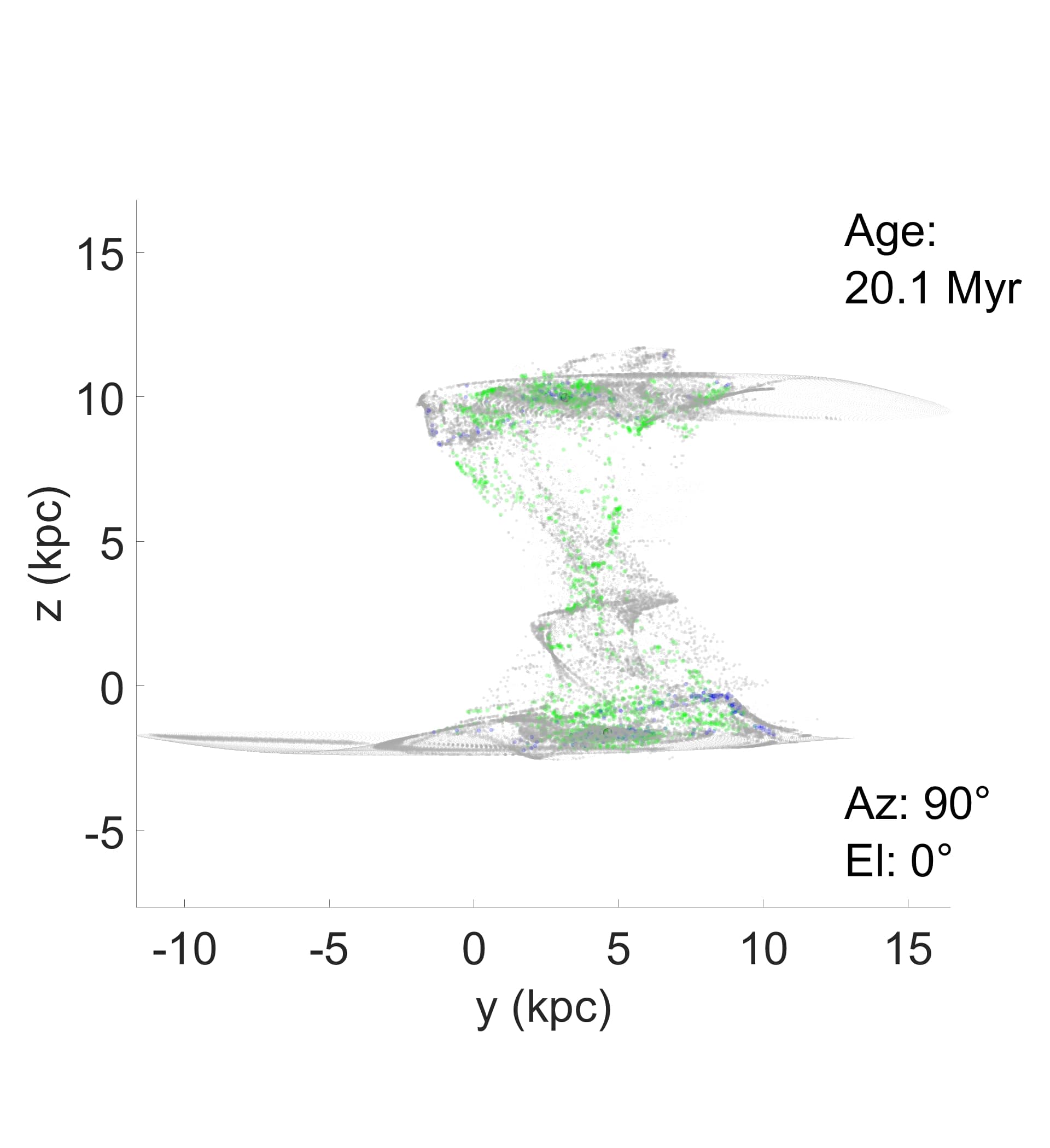}
        }
                \subfigure[]{%
          \includegraphics[width=0.45\textwidth]{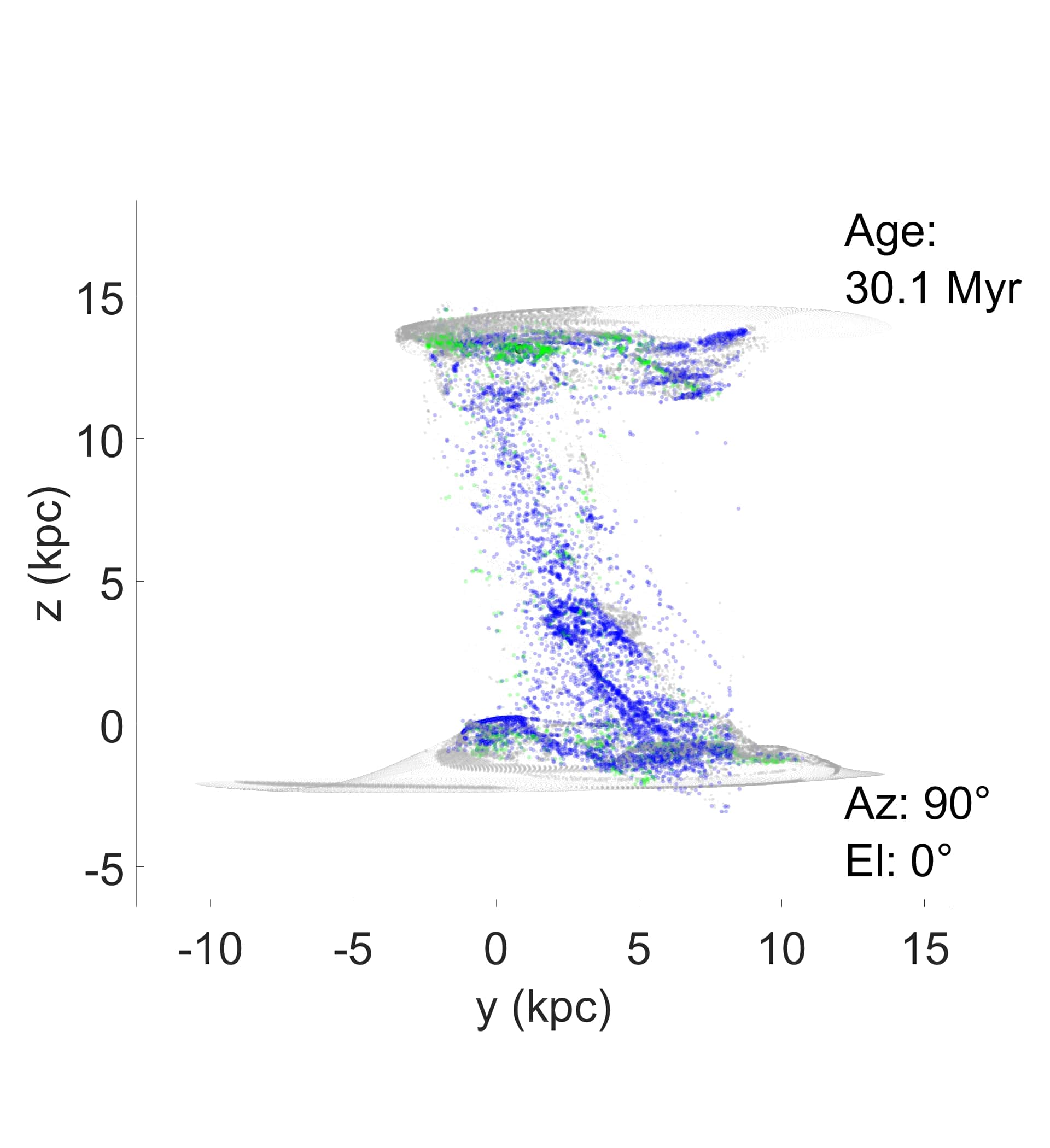}
        }\\
\caption{Panels (a) and (b) illustrate a 500 pc impact offset, 20\si{\degree} inclination disk-disk collision.  Panels (c) and (d) are of a 10 kpc offset, 20\si{\degree} disk-disk collision.  In blue are gas clouds that have collapsed less than 10 Myr ago.  The blue marks the most likely locations of ultra-luminous stars.  In bright green are gas clouds that have collapsed over 10 Myr ago but are within 10 Myr of a collision-induced gravitational instability.  Gray shows gas clouds that have either never collapsed, or that 10 Myr have elapsed since their last instability.} %
\label{fig:ostars1}
    \end{center}
\end{figure*}

\begin{figure*}
     \begin{center}
                \subfigure[]{%
          \includegraphics[width=0.45\textwidth]{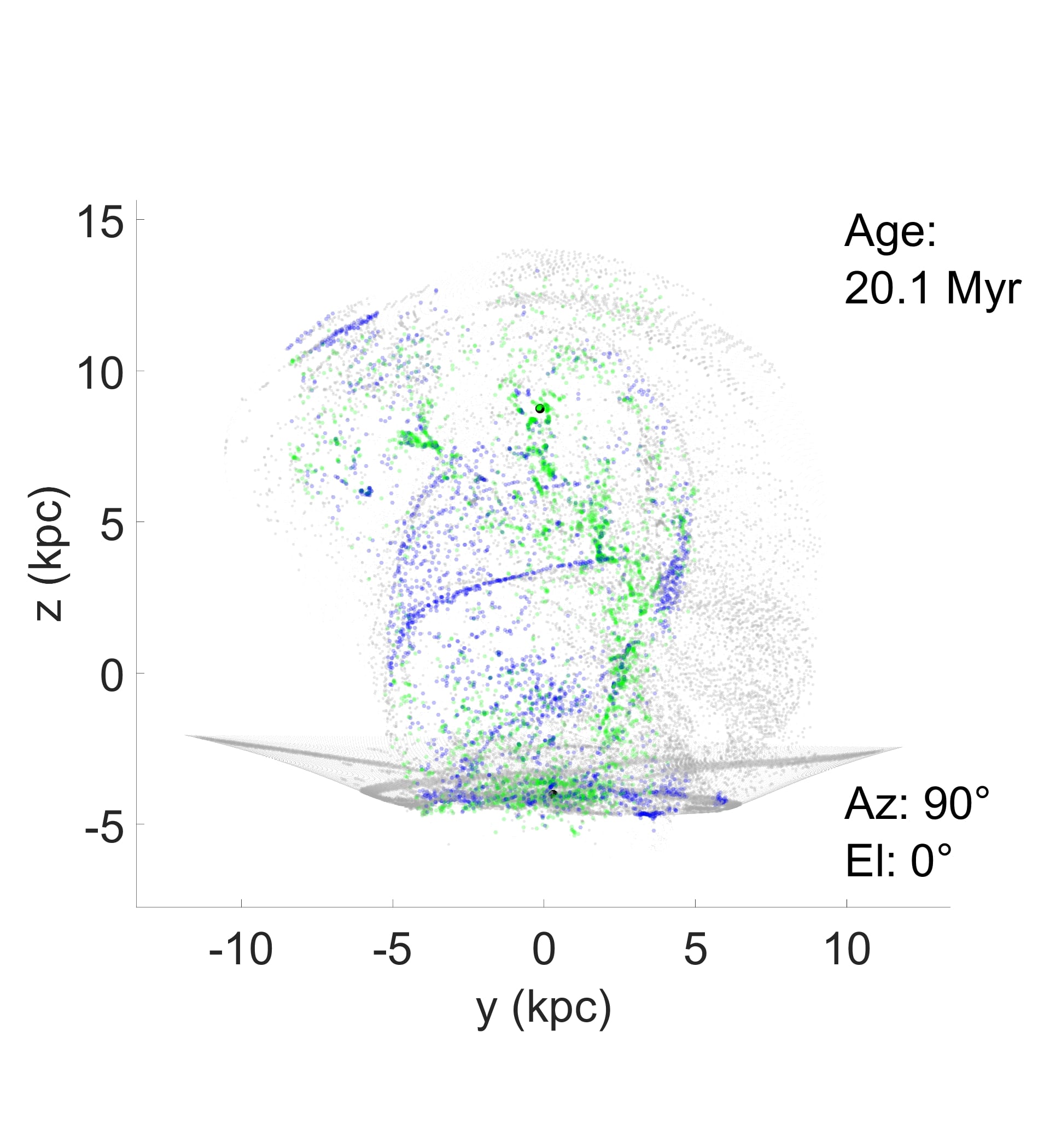}
        }
                \subfigure[]{%
          \includegraphics[width=0.45\textwidth]{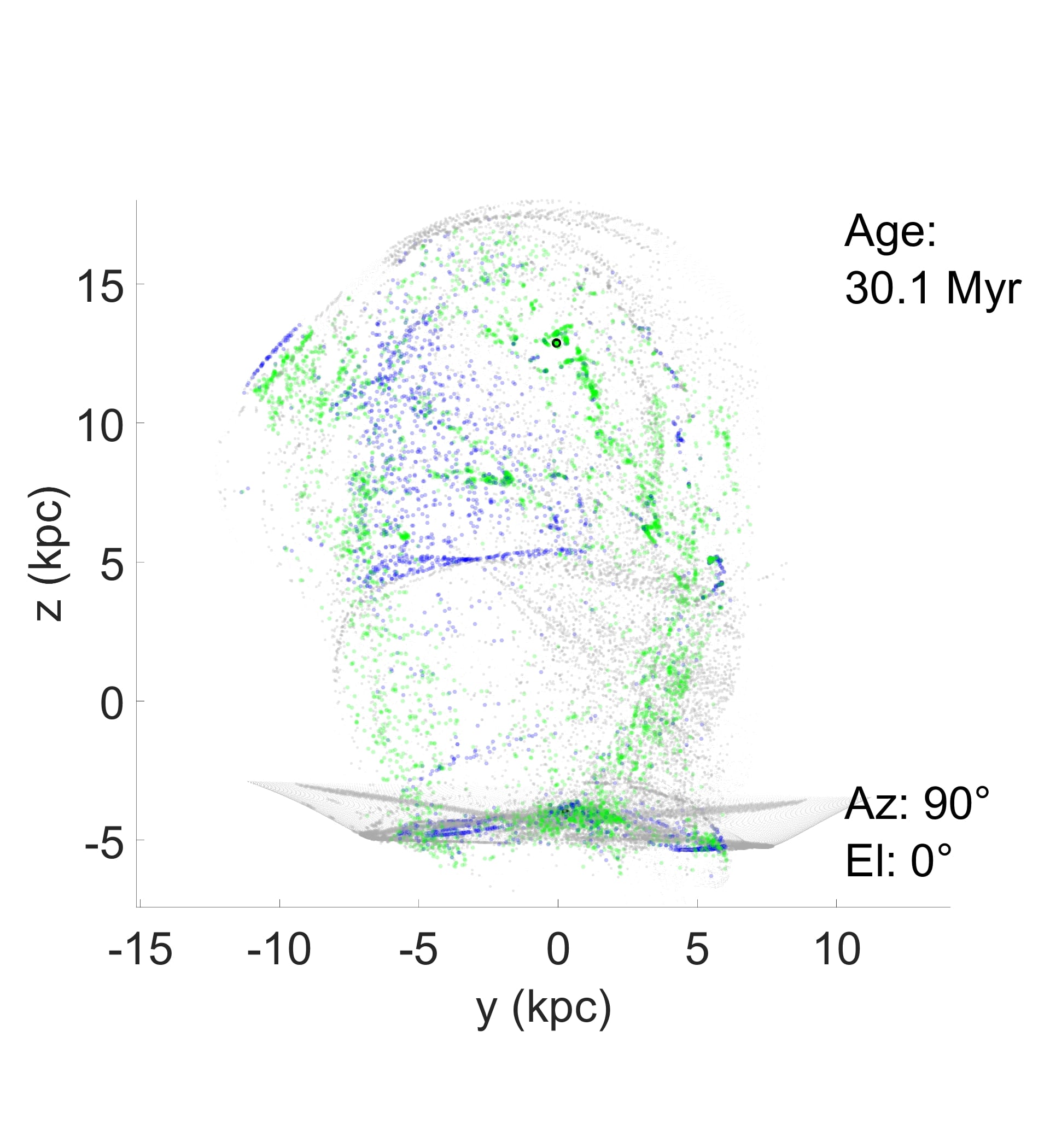}
        }\\ %  ------- End of the first row ----------------------%
        \subfigure[]{%
          \includegraphics[width=0.45\textwidth]{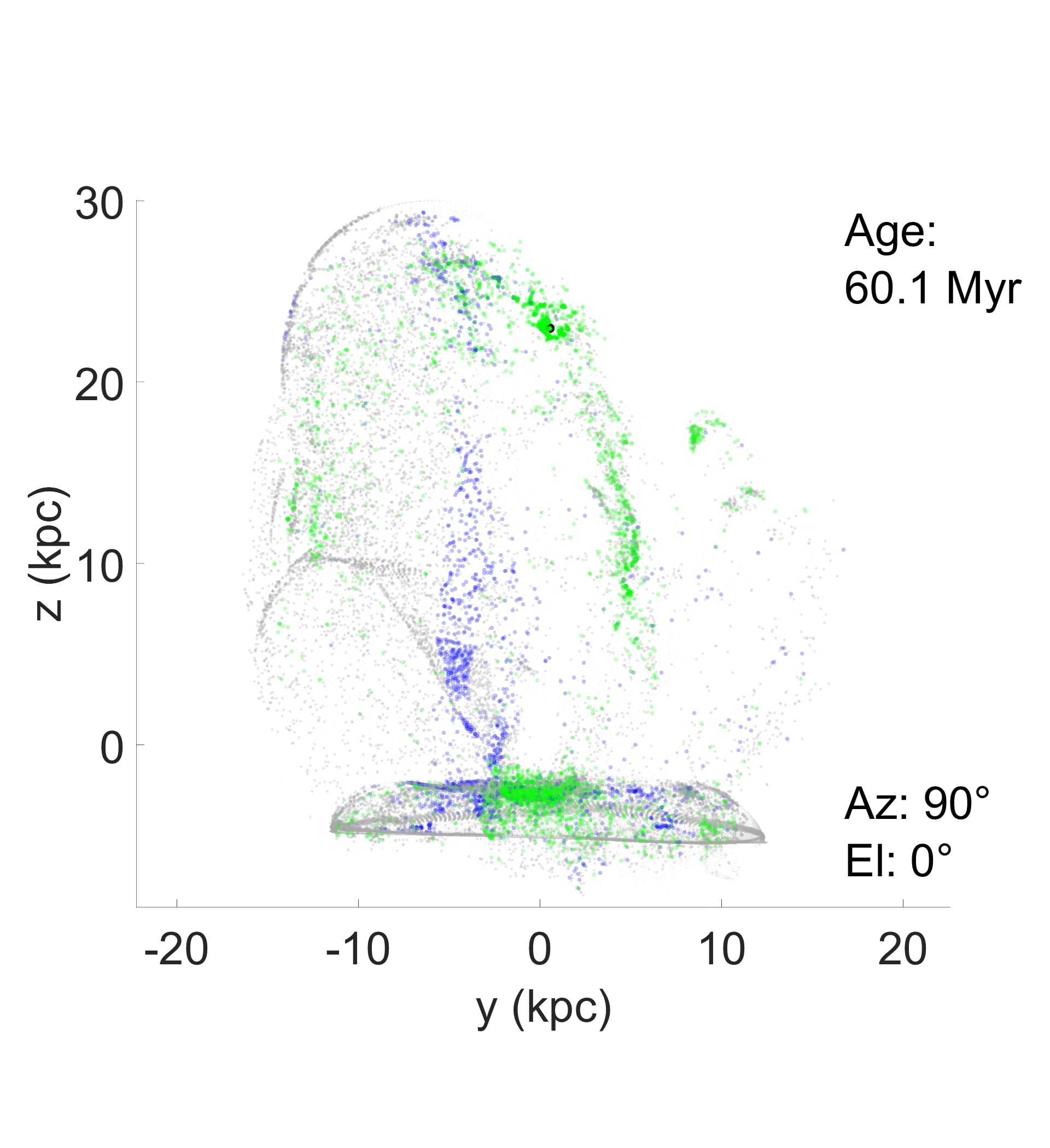}
        }
                \subfigure[]{%
          \includegraphics[width=0.45\textwidth]{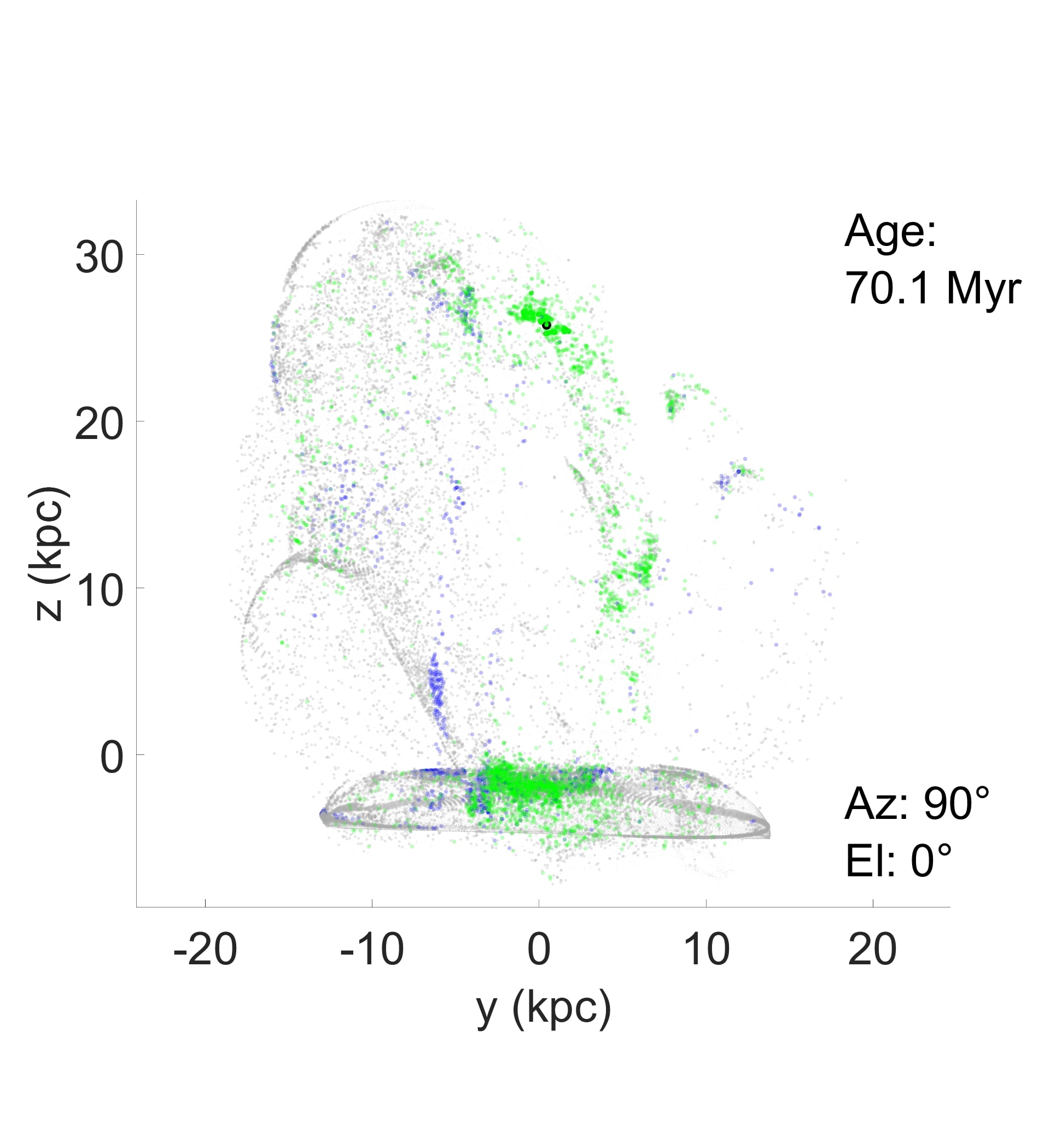}
        }\\
\caption{As \cref{fig:ostars1}, but for a 500 pc impact offset, 90\si{\degree} inclination disk-disk collision at various ages as given in the top right corner of each panel.  The colors are explained in \cref{fig:ostars1}.} %
\label{fig:ostars2}
    \end{center}
\end{figure*}

\indent To help see where these bridges may be illuminated by visible light we have created \Cref{fig:ostars1,fig:ostars2} that highlight where gas clouds have undergone an instability within the last 10 million years.  The ages of 20 Myr and 30 Myr shown in the figures are within the age estimates of the Taffy Galaxies.  This amount of time is long enough that the most luminous stars will have gone supernova.  The blue dots show the gas clouds that have completed a freefall time within the last 10 million years, and presumably formed stars.  The green color shows gas clouds that have initially collapsed more than 10 Myr ago, but have experienced a second cloud collision shock that has caused an instability within 10 Myr.  This second criteria of multiple collisions is a rough indicator of the sites of massive cloud buildup, and so, also for the sites of massive (e.g., O-type) star formation.  Our 100 pc resolution scale cannot resolve these processes directly, so we use this approximation.  

\indent We see for the 20\si{\degree} relative inclination disk-disk collision in \Cref{fig:ostars1} that there are a number of gas clouds shown in green that may contain massive stars.  The gas clouds in bright green in panel (a) and (c) at 20 Myr completed their first collapses within a few million years after the disk-disk collision.  The first wave of massive stars went supernova by around 10 Myr after the collision.  Around 26 Myr after the collision the splash bridge begins to light up with very luminous stars again.  By 30 Myr, as shown in panel b) and d), nearly the full span of the bridge are illuminated by recently collapsed gas clouds.  

\indent \Cref{fig:ostars2} shows a 90 \si{\degree} inclination, 500 pc impact offset disk-disk collision at various ages.  Massive star production histories of inclined disk-disk collisions are represented well by the nearly constant gas cloud collapse rates, though these are lower than the bursts of the previous case.  We predict there will be O-type stars present in the splash bridges of inclined collision continuously for the 70 Myr after the point of gas disk-disk contact.  Thus, inclined disks are predicted to remain luminous for long periods of time but, will be fainter at their peak luminosity than low-inclination splash bridges at the peak of their gas cloud collapse rate.

\flushbottom
\clearpage

\section{Conclusion}

\indent The models described above predict many regularities in the star formation histories resulting from direct gas-rich, disk-disk galaxy collisions.  For example, star formation can experience significant delays which are dependent on the collision parameters. The most important factor for determining time delays are the relative velocities of the disk-disk collisions.  The dependence of collision velocity is affirmed when comparing our models of 650-900 \kms{} collisions with 1000 \kms{} collisions.  The role of collision velocity can also be inferred from the dependence between shock heating strength and the corresponding cooling timescales.  The collision velocity directly affects the amount of thermal energy injected into the gas.  The thermal energy produced from shock waves can prevent gravitational instability in some gas clouds for tens of millions of years.

\indent The relative inclination of the disk-disk collision and the offset between the disks play a secondary role in the star formation delays. The offset distance between the centers of each galaxy disk has the obvious role in determining the amount of gas that will directly interact.  Large offsets result in only the outer regions experiencing strong shock waves.  Low-offset collisions paired with a low relative inclination will produce galaxy-wide shocks.

\indent The relative inclination between the two galactic disks is very important in determining the maximum intensity and burst-like character of star formation over time.  Low inclinations ($< 20\degree$) result in two strong bursts of star formation.  The gas clouds across both galaxies undergo shocks at approximately the same time, leading to star bursts.  Low inclinations will see a first burst of star formation 1-3 Myr after the disk-disk collision followed by a second burst of stars approximately 30 Myr later.  The observational age estimates of the Taffy Galaxies (UGC 12914/5) place this system right before the predicted second starburst.  The first wave of star formation 20-25 Myr ago, predicted by the models, would leave many low mass stars scattered throughout the bridge. This corroborates the findings from \citet{gao03}, whose observations suggest that star formation could be causing a portion of the radio continuum emission.   

\indent Highly inclined disk-disk collisions ($> 45\degree$) do not produce intense bursts of star formation in our models.  The geometry of these collisions creates an approximately 10 Myr time interval during which disk-disk interaction continues.  The geometry of this type of collision prevents the number of concurrently shining massive stars from producing the same peak emission levels as low-inclination collisions.  The star formation rate for highly inclined disk collisions ramps up over 10 Myr before then remaining fairly constant for the rest of the simulation time, approximately 120 Myr after the disk-disk collision.

These results can be compared to those of the studies of tidal bridges and tails, including those cited in the introduction. Firstly, we note that the delayed starbursts in low-inclination collisions are evidently unique among tidal structures, and a uniquely deterministic result of the cooling following a particular type of cloud-cloud collision that occurs nearly simultaneously across the galaxy disks. Secondly, the more constant bridge star formation rates in highly inclined collisions are qualitatively more similar to the star formation histories of tidal structures. Indeed, those with low-impact offsets seem to have modest peaks that recall a similar result found by Mullan et al. (2013) in their tail sample, though the timescales are much shorter in the present case. The high-inclination bridges are not purely impact products, but have a partial tidal character, and so are more complex.

Thirdly, there may be some similarities between the central bridge structures formed in low-inclination collisions, or the large clumps formed in some high-inclination collisions, and tidal dwarf galaxies. All seem to result from some kind of pile-ups, though the details might be quite different in the different cases.

Fourth, as implied by Figs. 5 and 6 the amount of star formation in the bridge is often nearly comparable to that in the disks. This contrasts with the order of magnitude smaller star formation rates in tidal tails relative to their parent galaxies in the sample of Mullan et al. (2013). This is not completely surprising because of the large gas fractions pulled out of the disks in the splash bridges, which is not generally the case in tidal tails. It is also interesting that most of the collapsing clouds in our models are involved in multiple cloud collisions, suggesting local gas pileups that would produce very massive clouds, and very massive star clusters capable of producing numerous massive stars. This may again be partially the result of the availability of large amounts of gas.

These comments are only a first exploration of possible comparisons between tidal structures and splash bridges, but they suggest directions for further work with both models and observations. In the coming decade, with greater sensitivity and resolution, new observations could provide a wealth of new information, challenging models to explain the complex physical processes.

\acknowledgments

\indent This work was performed under the auspices of the U.S. Department of Energy by Lawrence Livermore National Laboratory under Contract DE-AC52-07NA27344.

%\appendix

%\section{Appendix information}

%Bibliography
\bibliography{travslibrary}{}
\bibliographystyle{aasjournal}

\end{document}